\documentclass[aps,prd,reprint,superscriptaddress]{revtex4-2}

\usepackage{graphicx}
\usepackage{dcolumn}
\usepackage{bm}
\usepackage{amsmath}
\usepackage{amssymb}
\usepackage{microtype}
\usepackage{float}
\usepackage{multirow}
\usepackage{mathrsfs}
\usepackage[title]{appendix}
\usepackage{xcolor}
\usepackage{comment}
\usepackage{booktabs}
\usepackage{mathtools}
\usepackage{anyfontsize}
\usepackage{textcomp}
\usepackage[caption=false]{subfig}
\usepackage{hyperref}
\usepackage[nameinlink,noabbrev]{cleveref}
\hypersetup{
    colorlinks = true,
    citecolor  = red,    
    linkcolor  = blue,   
    urlcolor   = blue    
}

\DeclareUnicodeCharacter{2212}{\textminus}

\begin{document}
\title{Gravitational Production of Non-minimally Coupled Vector Dark Matter in Warm Inflation}

\author{Phongsakorn Sereewat}
\email{phongsakorn.ser@student.mahidol.edu}
\affiliation{Department of Physics, Faculty of Science, Mahidol University, 272 Rama VI Road, Ratchathewi, Bangkok 10400, Thailand}

\begin{abstract}
We study the purely gravitational production of a massive vector dark matter with a
non-minimal coupling to the Einstein tensor, $\xi G_{\mu\nu}X^\mu X^\nu$, in warm inflation.
The background, $V=\lambda\phi^4$ with a dissipation coefficient linear in temperature, is
constrained by CMB observations and parametrized by the dissipation ratio
$10^{-4}\le Q_0\le1$. We consider both unavoidable production channels: freeze-in from the
Standard Model bath through graviton exchange, and cosmological gravitational particle
production (CGPP). For freeze-in, we compute the scattering amplitudes including the contact
interaction induced by $\xi$, which enhances the production of light dark matter through the
longitudinal mode and suppresses that of heavy dark matter through destructive interference.
For CGPP, the slow-roll parameter is not small during warm inflation, which enhances the
production of light dark matter by up to three orders of magnitude relative to cold inflation.
Combining both channels, CGPP excludes most of the light freeze-in branch, leaving
$m_X\lesssim2.4\times10^{-5}$~GeV, a narrow window around $10^{10}$~GeV for strong dissipation,
and $m_X\gtrsim10^{14}$--$10^{15}$~GeV. Only the heavy branch remains viable for all
$|\xi|\lesssim1$.
\end{abstract}
\maketitle
\tableofcontents
\section{Introduction}
Cosmic inflation~\cite{PhysRevD.23.347}, a period of accelerated expansion in the early
Universe, provides the leading explanation for the observed homogeneity and isotropy of our
Universe. How inflation connects to the Standard Model (SM), however, remains an open
question. In the conventional picture, the inflaton oscillates around the minimum of its
potential after inflation ends and decays into SM particles, which then thermalize to form
the SM radiation bath. We refer to this scenario as cold inflation. By contrast,
Refs.~\cite{Bastero_Gil_2016, Berghaus_2025} showed, within a consistent effective field
theory, that a thermal bath can be sustained throughout inflation itself. In this case the
inflaton continuously dissipates its energy into SM radiation during inflation, and the
Universe enters the radiation-dominated epoch smoothly, without a separate reheating stage
driven by inflaton oscillations. This scenario is known as warm inflation; see
Ref.~\cite{Berera_2009} for a review.

Because warm inflation maintains a hot SM bath during inflation, dark matter (DM) can be
produced efficiently by scattering of SM particles in this bath. Such production also occurs
in cold inflation, but only after the bath forms during reheating, and is therefore expected
to be considerably more efficient in warm inflation. Scalar DM produced in this way has been
studied in Ref.~\cite{Wang_2025}, whereas vector DM remains unexplored. Vector DM is also less constrained than a light scalar by the primordial power spectrum:
a light scalar spectator acquires isocurvature perturbations in conflict with the CMB,
whereas the vector power spectrum is suppressed on large scales and its long-wavelength
perturbations are adiabatic~\cite{Graham:2015rva}. Purely gravitational production of vector
DM has been studied in cold inflation~\cite{Ema:2019yrd, Ahmed:2020fhc, Kolb:2020fwh}, but
not in warm inflation.

Thermal scattering is not the only source of dark matter, however. Even in the absence of
any coupling to the SM, the expansion of the Universe excites DM quanta out of the vacuum,
a process known as cosmological gravitational particle
production (CGPP; see
Ref.~\cite{kolb2025cosmologicalgravitationalparticleproduction} for a review). Purely
gravitational DM therefore has two unavoidable production channels: freeze-in from the SM
bath through graviton exchange, and vacuum excitation by the expanding background.

Moreover, the minimal Proca theory is not the most general description of a massive
vector in curved spacetime. Effective field theory allows, and generically requires,
non-minimal couplings of $X_\mu$ to curvature. Such terms are consistent with the symmetries
of the Proca theory, and even if absent at tree level they are generated once
self-interactions or loop corrections are included~\cite{Capanelli:2024pzd}. Among them,
the coupling to the Einstein tensor, $\xi\,G_{\mu\nu}X^\mu X^\nu$, belongs to the
generalized Proca theories~\cite{Heisenberg:2014rta, Tasinato:2014eka} and is the
unique quadratic coupling that avoids Ostrogradsky instabilities~\cite{Ozsoy:2023gnl}. This
coupling affects both production channels. It modifies the dispersion relations of the
vector modes on the inflationary background, which changes cosmological gravitational
particle production~\cite{Capanelli:2024rlk, Cembranos:2023qph}. It also adds a contact
interaction with the SM energy-momentum tensor, which enhances freeze-in from the thermal
bath through the longitudinal mode~\cite{Barman:2021qds}. The non-minimal case therefore
tests how robust the minimal results are, and how strongly the viable DM mass range depends
on the gravitational couplings of the dark sector.

The organization of this work is as follows. In Sec.~\ref{sec:WI} we review the warm-inflation
background and identify the benchmarks allowed by CMB observations. In Sec.~\ref{sec:grav} we
compute the gravitational freeze-in of $X$ from the SM bath, including the non-minimal coupling
$\xi$ in the scattering amplitudes, the production rates and the relic abundance. In
Sec.~\ref{sec:CGPP} we study the cosmological gravitational particle production of $X$ in the
warm-inflation background, derive the stability conditions for $\xi\neq0$, and combine both
channels to constrain the $(m_X,\xi)$ plane. We conclude in Sec.~\ref{sec:conclusion}.

\section{Warm Inflation Background}
\label{sec:WI}

In warm inflation (WI), the inflaton dissipates its energy into a radiation bath
while inflation is still ongoing, so the background evolution is governed jointly
by the two components. Their continuity equations read
\begin{align}
    \dot{\rho}_\phi + 3H(\rho_\phi + p_\phi) &= -\Upsilon(\phi,T)\,(\rho_\phi + p_\phi), \label{eq:cont_phi}\\
    \dot{\rho}_r + 3H(\rho_r + p_r) &= +\Upsilon(\phi,T)\,(\rho_\phi + p_\phi), \label{eq:cont_r}
\end{align}
where the subscripts $\phi$ and $r$ denote the inflaton and radiation, respectively,
and the dissipation coefficient $\Upsilon$ (with mass dimension one) sets the rate of
energy transfer. For a homogeneous inflaton,
\begin{equation}
    \rho_\phi = \frac{1}{2}\dot{\phi}^2 + V(\phi), \qquad
    p_\phi = \frac{1}{2}\dot{\phi}^2 - V(\phi),
    \label{eq:rho_p}
\end{equation}
so that $\rho_\phi + p_\phi = \dot{\phi}^2$. Substituting Eq.~\eqref{eq:rho_p} into
Eqs.~\eqref{eq:cont_phi} and \eqref{eq:cont_r}, and using $p_r = \rho_r/3$, we obtain
\begin{align}
    \ddot{\phi} + (3H + \Upsilon)\dot{\phi} + V'(\phi) &= 0, \label{eq:eom_phi}\\
    \dot{\rho}_r + 4H\rho_r &= \Upsilon\dot{\phi}^2. \label{eq:eom_r}
\end{align}
Together with the Friedmann equation,
\begin{equation}
    H^2 = \frac{\rho_\phi + \rho_r}{3M_P^2},
    \label{eq:friedmann}
\end{equation}
these equations determine the background dynamics of WI, where $M_P$ is the reduced
Planck mass. Throughout this work, we assume that the energy density of the dark matter
(DM), $\rho_X$, satisfies $\rho_X \ll \rho_r+\rho_\phi$, so that its backreaction on the background
evolution can be safely neglected.

It is convenient to introduce the dissipation ratio $Q \equiv \Upsilon/(3H)$;
$Q < 1$ and $Q > 1$ correspond to the weak and strong dissipative regimes. During slow
roll, radiation production balances its redshift, $\dot{\rho}_r \simeq 0$, so that
\begin{equation}
    \rho_r \simeq \frac{\Upsilon\dot{\phi}^2}{4H} = \frac{3}{4}\,Q\,\dot{\phi}^2 .
    \label{eq:rho_r_sr}
\end{equation}

The temperature of the radiation bath is
\begin{equation}
    T = \left(\frac{30\,\rho_r}{\pi^2 g_*}\right)^{1/4},
    \label{eq:T}
\end{equation}
where $g_*$ is the effective number of relativistic degrees of freedom. Above the
electroweak scale the Standard Model (SM) gives $g_* = 106.75$; if the inflaton quanta
are also thermalized, they contribute one additional bosonic degree of freedom,
$g_* = 107.75$. A consistent WI regime requires $T > H$, so that thermal effects
dominate over de Sitter vacuum fluctuations.

Determining $\Upsilon(\phi,T)$ from first principles requires thermal field theory. A dissipation
coefficient $\Upsilon=C_T\,T$ together with a quartic potential $V(\phi)=\lambda\phi^4$ has been
shown to realize successful warm inflation~\cite{Bastero_Gil_2016, Berghaus_2025}. We therefore
adopt this background for the remainder of this work. We expect the key qualitative features of
our results to hold for other models in which dissipation sustains $T\gg H$.

In warm inflation the primordial curvature perturbation is sourced not only by the quantum
fluctuations of the inflaton but also by the thermal noise of the radiation bath. Assuming that
the inflaton fluctuations are thermalized, the scalar power spectrum at horizon crossing
reads~\cite{Bartrum:2013fia,Ramos:2013nsa}
\begin{equation}
\Delta_\mathcal{R}^2=\left(\frac{H^2}{2\pi\dot\phi}\right)^2
\left[1+2n_{\rm BE}+\frac{2\sqrt3\,\pi Q}{\sqrt{3+4\pi Q}}\,\frac{T}{H}\right]G(Q),
\label{eq:DeltaR}
\end{equation}
where $n_{\rm BE}=(e^{H/T}-1)^{-1}$ is the Bose--Einstein occupation of the inflaton modes and
$Q=\Upsilon/3H$. The first term in the bracket is the vacuum contribution of cold inflation, the
second its thermal enhancement, and the third the noise associated with dissipation. The function
$G(Q)$ accounts for the growth of the inflaton perturbations through their coupling to the
radiation fluctuations when $\Upsilon$ depends on temperature~\cite{Graham:2009bf}; for
$\Upsilon\propto T$ we use the numerical fit
$G(Q)=1+0.335\,Q^{1.364}+0.0185\,Q^{2.315}$~\cite{Benetti:2016jhf}. The spectral index and the
tensor-to-scalar ratio follow as
\begin{equation}
n_s-1=\frac{\mathrm{d}\ln\Delta_\mathcal{R}^2}{\mathrm{d}\ln k},\qquad
r=\frac{2H^2}{\pi^2M_P^2\,\Delta_\mathcal{R}^2}.
\end{equation}
We evaluate these quantities at the pivot scale, which leaves the horizon 60 e-folds before the
end of inflation, and parametrize the models by $Q_0$, the value of $Q$ two e-folds before the
pivot. 

\begin{figure}
  \includegraphics[width=0.8\columnwidth]{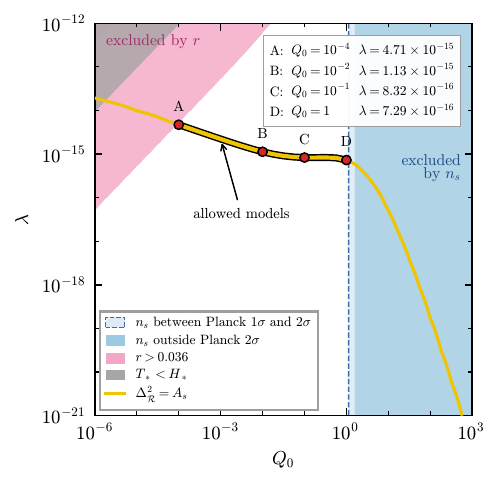}
  \caption{CMB constraints on warm inflation with $V=\lambda\phi^4$ and $\Upsilon=C_T T$.
  Physical models lie on the yellow line $\Delta_\mathcal{R}^2=A_s$. Shaded regions are excluded
  by $n_s$ (Planck $2\sigma$), $r>0.036$, and $T_*<H_*$. The outlined segment is allowed;
  A--D mark the benchmark models.}
  \label{fig:cmb_constraints}
\end{figure}

For each $Q_0$, the self-coupling $\lambda$ is then fixed by the amplitude measured by
\textit{Planck}, $\Delta_\mathcal{R}^2=A_s\simeq2.1\times10^{-9}$, and the model is required to satisfy
$n_s=0.9649\pm0.0042$~\cite{Planck:2018jri} and $r<0.036$~\cite{BICEP:2021xfz}, as well as
$T>H$. The resulting constraints are shown in Fig.~\ref{fig:cmb_constraints}.

\begin{figure}
  \includegraphics[width=0.8\columnwidth]{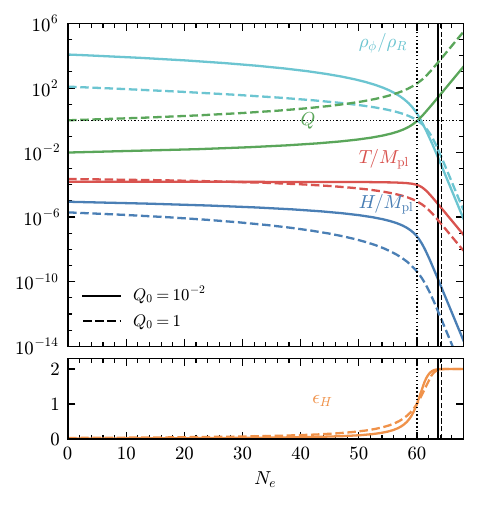}
  \caption{Background evolution in warm inflation with $V=\lambda\phi^4$ and $\Upsilon=C_T T$, for $Q_0=10^{-2}$, $\lambda=1.13\times10^{-15}$ (solid; $n_s=0.966$, $r=7.5\times10^{-3}$) and $Q_0=1$, $\lambda=7.29\times10^{-16}$ (dashed; $n_s=0.968$, $r=3.2\times10^{-4}$). Inflation ends when $\epsilon_H=1$, 60 e-folds after the pivot scale exits the horizon, and radiation domination begins when $\epsilon_H\simeq2$.}
  \label{fig:wi_background}
\end{figure}

From the CMB-allowed models in Fig.~\ref{fig:cmb_constraints}, we choose two benchmarks: a weak
dissipative regime, $Q_0=10^{-2}$, and a strong one, $Q_0=1$. Their background evolution is
shown in Fig.~\ref{fig:wi_background}. The dissipation ratio $Q\equiv\Upsilon/3H$ measures how
efficiently the inflaton loses energy to the SM radiation bath compared with Hubble friction.
In both cases $Q$ grows monotonically during and after inflation. As a result, the ratio $\rho_\phi/\rho_r$
decreases steadily and reaches unity close to the end of inflation, after which radiation
dominates. Warm inflation therefore connects the quasi-de Sitter phase smoothly to the
radiation-dominated epoch, without a separate reheating stage. During the last e-folds of
inflation the temperature decreases only slowly, with $T_e\simeq10^{-5}$--$10^{-4}\,M_P$ at the
end of inflation, far above the electroweak scale; for $Q_0=1$, $T$ is about nine times larger
fifteen e-folds before the end. The Hubble rate is nearly constant, since the potential energy
dominates. After inflation both fall rapidly, with $T\propto a^{-1}$ and $H\propto a^{-2}$ in the
radiation era.

\section{Thermal Production}
\label{sec:grav}

We start from the action
\begin{equation}
    S = \int \mathrm{d}^4x\, \sqrt{-g}\left[ \frac{M_P^2}{2}R + \mathcal{L}_M \right],
    \label{eq:action}
\end{equation}
where $\mathcal{L}_M$ contains the inflaton, the SM fields, and the
DM. We take the DM to be a massive Abelian vector (Proca) field $X_\mu$
that couples to the other sectors only through gravity,
\begin{align}
    \mathcal{L}_X &= -\frac{1}{4} X^{\mu\nu}X_{\mu\nu}
                    - \frac{1}{2} m_X^2\, X^\nu X_\nu + \xi G_{\mu\nu}X^\mu X^{\nu}  ,\label{eq:proca}
    \\ X_{\mu\nu} &= \partial_\mu X_\nu - \partial_\nu X_\mu ,
\end{align}
with metric signature $(-,+,+,+)$, where $G_{\mu\nu}$ is the Einstein 
tensor and $\xi$ is a dimensionless non-minimal coupling. This model 
belongs to the class of \textit{generalized Proca} 
theories~\cite{Heisenberg:2014rta, Tasinato:2014eka, BeltranJimenez:2016rff}. 
Gravitational production of vector DM, including non-minimal couplings to
curvature, has been studied for cold inflation followed by a separate
reheating stage~\cite{Ozsoy:2023gnl, Cembranos:2023qph,
Capanelli:2024pzd, Capanelli:2024rlk}, where such couplings were
found to induce gradient, ghost, or runaway instabilities. To our knowledge,
gravitational production of vector DM in warm inflation, where a radiation
bath is present throughout inflation and the transition to radiation
domination proceeds smoothly, remains unexplored.

Throughout this work we assume the weak-gravity regime: all relevant energy scales,
including the bath temperature and the Hubble rate, lie far below the Planck scale,
$T, H \ll M_P$. We therefore split the metric into a classical
Friedmann--Lema\^{i}tre--Robertson--Walker (FLRW) background $\bar g_{\mu\nu}$ and a
small quantum fluctuation $h_{\mu\nu}$,
\begin{equation}
    g_{\mu\nu} = \bar g_{\mu\nu} + \frac{2}{M_P}\, h_{\mu\nu},
    \qquad |h_{\mu\nu}| \ll M_P ,
    \label{eq:metric_split}
\end{equation}
and expand the action to leading order in $1/M_P$:
\begin{multline}
    S = \int \mathrm{d}^4x\,\sqrt{-\bar g}\,\bigg[ \frac{M_P^2}{2}\bar R + \bar{\mathcal{L}}_M + \mathcal{L}_2(h) \\
      + \frac{1}{M_P}\Big( h_{\mu\nu} T_M^{\mu\nu} + \mathcal{L}_3(h) \Big) + \dots \bigg],
\label{eq:action_expanded}
\end{multline}
where $\bar{\mathcal{L}}_M \equiv \mathcal{L}_M(\bar g)$ denotes the matter Lagrangian on the
FLRW background, $\mathcal{L}_n(h)$ is the term of $n$-th order in $h_{\mu\nu}$, and
$T_M^{\mu\nu}$ is the energy-momentum tensor of the matter fields, including the contribution
of the non-minimal coupling. Quantizing
$h_{\mu\nu}$ perturbatively, we identify it with the graviton, and $h_{\mu\nu}T_M^{\mu\nu}$
describes the leading-order graviton--matter interaction.

The expanded action in Eq.~\eqref{eq:action_expanded} naturally separates DM production
into two channels: (i) cosmological gravitational particle production, arising
from the DM part of $\bar{\mathcal{L}}_M$, in which the DM field propagates freely on the
time-dependent FLRW background; and (ii) freeze-in through $s$-channel graviton-mediated
scattering of bath particles, arising from the interaction term $h_{\mu\nu}T_M^{\mu\nu}/M_P$.

We first focus on DM production via graviton-mediated annihilation. This process entails
a two-step calculation. The first step requires the computation of the annihilation cross
section. Following this, the Boltzmann equation is solved using the calculated cross
section to determine the relic abundance of DM.

\subsection{Graviton exchange}
\label{sec:freezein}

To compute the scattering amplitudes we neglect spacetime curvature and the cosmic
expansion during each scattering event. This is justified because the duration of a
collision, $\sim 1/\sqrt{s}\sim 1/T$, is much shorter than the Hubble time $1/H$
throughout and after WI, where $T\gg H$. We may therefore evaluate the energy-momentum
tensors on the Minkowski metric $\eta_{\mu\nu}={\rm diag}(-1,+1,+1,+1)$. For a real
scalar $\varphi$, a Dirac fermion $\psi$, and a massive vector $A_\mu$, the energy-momentum tensors read
\begin{align}
T^{\mu\nu}_{\varphi} &= \partial^\mu\varphi\,\partial^\nu\varphi
  - \frac{1}{2}\eta^{\mu\nu}\left(\partial_\alpha\varphi\,\partial^\alpha\varphi
  + m_\varphi^2\varphi^2\right), \label{eq:T_scalar}\\
T^{\mu\nu}_{\psi} &= -\frac{i}{4}\,\bar\psi\left(\gamma^\mu \overleftrightarrow{\partial^\nu}
  + \gamma^\nu \overleftrightarrow{\partial^\mu}\right)\psi \nonumber\\
  &\quad + \eta^{\mu\nu}\,\bar\psi\left(i\gamma^\alpha\partial_\alpha - m_\psi\right)\psi,
  \label{eq:T_fermion}\\
T^{\mu\nu}_{A} &= F^{\mu\alpha}F^{\nu}{}_{\alpha}
  - \frac{1}{4}\eta^{\mu\nu}F_{\alpha\beta}F^{\alpha\beta} \nonumber\\
  &\quad + m_A^2\Big(A^\mu A^\nu - \frac{1}{2}\eta^{\mu\nu}A_\alpha A^\alpha\Big),
  \label{eq:T_vector}
\end{align}
where $\bar\psi\overleftrightarrow{\partial}\psi \equiv \bar\psi(\partial\psi)
- (\partial\bar\psi)\psi$, the gamma matrices satisfy
$\{\gamma^\mu,\gamma^\nu\} = -2\eta^{\mu\nu}$, $F_{\mu\nu}=\partial_\mu A_\nu-\partial_\nu A_\mu$, and
$F^{\nu}{}_{\alpha}=F^{\nu\beta}\eta_{\beta\alpha}$. All three tensors are symmetric and
conserved on shell, and the bosonic ones have positive energy density $T^{00}$. The second
term in Eq.~\eqref{eq:T_fermion} vanishes on shell and does not contribute to the
amplitudes. The DM tensor $T^{\mu\nu}_X$ follows from Eq.~\eqref{eq:T_vector} with
$A_\mu\to X_\mu$ and $m_A\to m_X$. For massless SM gauge bosons we use
Eq.~\eqref{eq:T_vector} with $m_A=0$ and only the two transverse polarizations.

\begin{figure}
    \centering
    \includegraphics[width=0.8\linewidth]{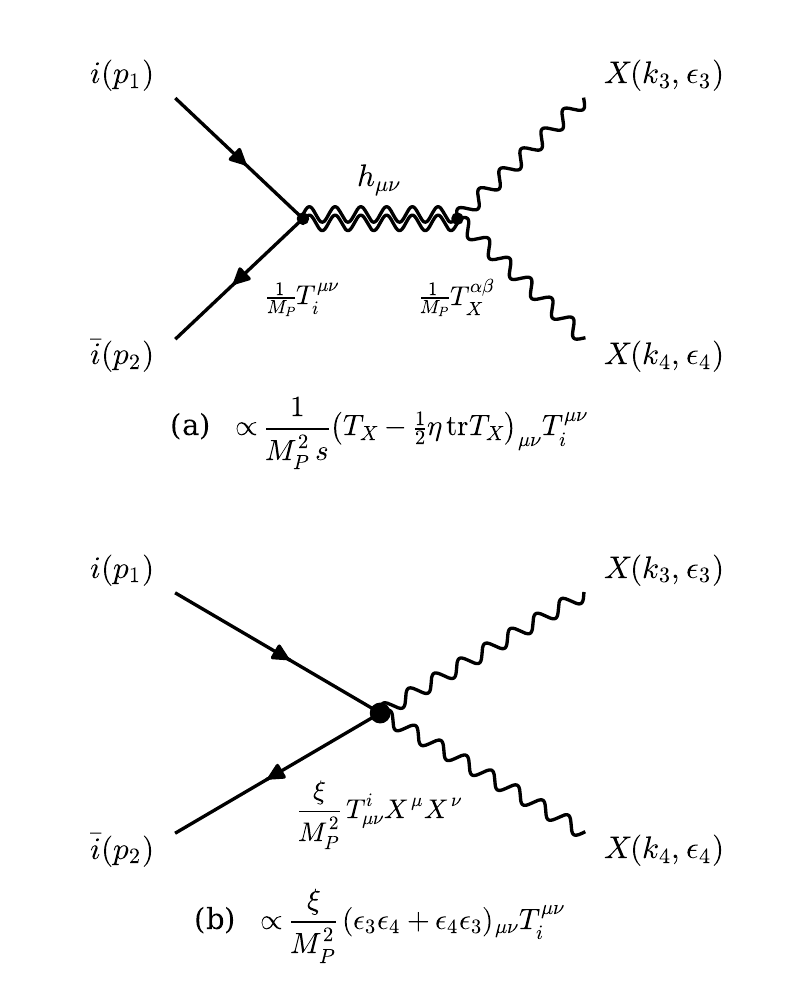}
    \caption{Graviton-mediated production of vector dark matter from the SM thermal bath,
    $i\,\bar{i}\to XX$, where $i$ denotes an SM scalar, fermion or gauge boson.
    (a)~$s$-channel graviton exchange: the graviton $h_{\mu\nu}$ couples to the
    energy-momentum tensors $T_i^{\mu\nu}$ and $T_X^{\alpha\beta}$ with strength $1/M_P$.
    (b)~Contact interaction induced by the non-minimal coupling
    $\xi G_{\mu\nu}X^\mu X^\nu$; since $G_{\mu\nu}=T^i_{\mu\nu}/M_P^2$ on the graviton
    sourced by the bath, the graviton pole cancels, leaving the local vertex
    $(\xi/M_P^2)\,T^i_{\mu\nu}X^\mu X^\nu$.}
    \label{fig:feynman}
\end{figure}

Since gravity couples to each species only through its own energy-momentum tensor,
there is no direct vertex connecting the bath particles to the DM. At leading order the
process $i\,i\to X X$ proceeds through $s$-channel graviton exchange and, for $\xi\neq0$,
through the contact interaction obtained by evaluating $\xi G_{\mu\nu}X^\mu X^\nu$ on the
graviton sourced by the bath, as shown in Fig.~\ref{fig:feynman}.
In the harmonic gauge the graviton propagator is
\begin{equation}
\Pi_{\mu\nu\rho\sigma}(k) = \frac{\eta_{\mu\rho}\eta_{\nu\sigma}
  + \eta_{\mu\sigma}\eta_{\nu\rho} - \eta_{\mu\nu}\eta_{\rho\sigma}}{2k^2},
\label{eq:propagator}
\end{equation}
and the amplitude takes the form
\begin{equation}
\mathcal{M}_{i\to X} = \frac{1}{M_P^2}\bigg[\frac{\mathcal{T}_i^{\mu\nu}\,
  \mathcal{T}^{X}_{\mu\nu} - \frac{1}{2}\,\mathcal{T}_i\,\mathcal{T}^{X}}{s}
  -\xi\,\mathcal{T}_i^{\mu\nu}\,\mathcal{C}_{\mu\nu}\bigg],
\label{eq:amplitude}
\end{equation}
where $\mathcal{T}_i^{\mu\nu}=\langle 0|T_i^{\mu\nu}|i\,i\rangle$ and
$\mathcal{T}^{X}_{\mu\nu}=\langle XX|T^X_{\mu\nu}|0\rangle$ are the on-shell vertex
factors, $\mathcal{C}_{\mu\nu}=\langle XX|X_\mu X_\nu|0\rangle
=\epsilon^*_{3\mu}\epsilon^*_{4\nu}+\epsilon^*_{4\mu}\epsilon^*_{3\nu}$, normalized such that
the mass term of $\mathcal{T}^X_{\mu\nu}$ is $m_X^2\,\mathcal{C}_{\mu\nu}$,
$\mathcal{T}\equiv\eta_{\mu\nu}\mathcal{T}^{\mu\nu}$ denotes the trace, and
$s=-(p_1+p_2)^2=-(p_3+p_4)^2$ is the squared center-of-mass energy. The first term is
graviton exchange, $\mathcal{M}_h$, and the second the contact term, $\xi\mathcal{M}_c$; their
relative sign is fixed by $G_{\mu\nu}=T_{\mu\nu}/M_P^2$, while an overall sign of $\mathcal{M}$
is irrelevant for what follows. We define the angle-integrated,
polarization-summed squared amplitude
\begin{equation}
\mathcal{A}_{i\to X} \equiv \int_{-1}^{1}\mathrm{d}\cos\theta
  \sum_{\rm pol}\left|\mathcal{M}_{i\to X}\right|^2 ,
\label{eq:Adef}
\end{equation}
where the sum runs over the polarizations of both initial and final states.

For a massive real vector DM $X_\mu$ in the final state, we obtain
\begin{align}
\mathcal{A}_{i\to X} &= \mathcal{A}^{(0)}_{i\to X} + \xi\,\mathcal{A}^{(1)}_{i\to X} + \xi^2\,\mathcal{A}^{(2)}_{i\to X}, \label{eq:A_xi_expansion}
\end{align}
which is exact at this order, since $\mathcal{M}$ is linear in $\xi$. The minimal terms are
\begin{align}
\mathcal{A}^{(0)}_{\varphi\to X} &= \frac{1}{M_P^4}\bigg[\frac{168\,m_\varphi^4 m_X^4}{15 s^2}
   + \frac{8 m_\varphi^2 m_X^2}{15 s}\big(17 m_\varphi^2 - 3 m_X^2\big) \nonumber\\
 &\quad + \frac{2}{15}\big(19 m_\varphi^4 - 44 m_\varphi^2 m_X^2 + 9 m_X^4\big) \nonumber\\
 &\quad + \frac{2s}{15}\big(3 m_X^2 - 7 m_\varphi^2\big) + \frac{s^2}{5}\bigg], \label{eq:A_phiX}\\
\mathcal{A}^{(0)}_{\psi\to X} &= \frac{1}{M_P^4}\bigg[\!-\frac{336\,m_\psi^4 m_X^4}{15 s^2}
   - \frac{4 m_\psi^2 m_X^2}{15 s}\!\big(68 m_\psi^2 + 3 m_X^2\big) \nonumber\\
 &\quad + \frac{4}{15}\big(6 m_X^4 - 11 m_\psi^2 m_X^2 - 19 m_\psi^4\big) \nonumber\\
 &\quad + \frac{7s}{15}\big(4 m_X^2 - m_\psi^2\big) + \frac{13 s^2}{30}\bigg], \label{eq:A_psiX}\\
\mathcal{A}^{(0)}_{A\to X} &= \frac{1}{M_P^4}\bigg[\frac{504\,m_A^4 m_X^4}{15 s^2}
   + \frac{136 m_A^2 m_X^2}{5 s}\big(m_A^2 + m_X^2\big) \nonumber\\
 &\quad + \frac{2}{15}\big(57 m_A^4 + 268 m_A^2 m_X^2 + 57 m_X^4\big) \nonumber\\
 &\quad + \frac{118 s}{15}\big(m_A^2 + m_X^2\big) + \frac{29 s^2}{15}\bigg], \label{eq:A_AX}\\
\mathcal{A}^{(0)}_{\gamma\to X} &= \frac{2\,(13 s^2 + 56 m_X^2 s + 48 m_X^4)}{15 M_P^4}. \label{eq:A_gamX}
\end{align}
the terms linear in $\xi$ are
\begin{align}
\mathcal{A}^{(1)}_{\varphi\to X} &= -\frac{2}{15M_P^4}\bigg[\Big(\frac{s}{m_X^2}+\frac{16m_X^2}{s}\Big)
   \big(s^2+2m_\varphi^2 s+6m_\varphi^4\big) \nonumber\\
 &\quad + 7s^2 - 76m_\varphi^2 s + 132 m_\varphi^4\bigg], \label{eq:A1_phiX}\\
\mathcal{A}^{(1)}_{\psi\to X} &= -\frac{s-4m_\psi^2}{15M_P^4}\bigg[\Big(\frac{s}{m_X^2}+\frac{16m_X^2}{s}\Big)
   \big(s+6m_\psi^2\big) \nonumber\\
 &\quad + 4\big(13s+33m_\psi^2\big)\bigg], \label{eq:A1_psiX}\\
\mathcal{A}^{(1)}_{A\to X} &= -\frac{2}{5M_P^4}\bigg[\Big(\frac{s}{m_X^2}+\frac{16m_X^2}{s}\Big)
   \big(s^2+2m_A^2 s+6m_A^4\big) \nonumber\\
 &\quad + 37s^2 + 164m_A^2 s + 132m_A^4\bigg], \label{eq:A1_AX}\\
\mathcal{A}^{(1)}_{\gamma\to X} &= -\frac{4s}{15M_P^4}\Big(\frac{s^2}{m_X^2}+52s+16m_X^2\Big), \label{eq:A1_gamX}
\end{align}
and the terms quadratic in $\xi$ are
\begin{align}
    \mathcal{A}^{(2)}_{\varphi\to X} &= \frac{1}{15M_P^4}\bigg[\frac{s^2}{m_X^4}\big(s^2+2m_\varphi^2 s+6m_\varphi^4\big) \nonumber\\
   &\quad+ \frac{2s}{m_X^2}\big(s-4m_\varphi^2\big)^2 \nonumber\\
 &\quad + 4\big(9s^2+8m_\varphi^2 s+64m_\varphi^4\big)\bigg], \label{eq:A2_phiX}\\
 \mathcal{A}^{(2)}_{\psi\to X} &= \frac{s-4m_\psi^2}{30M_P^4}\bigg[\frac{s^2}{m_X^4}\big(s+6m_\psi^2\big) \nonumber\\
   &\quad+ \frac{4s}{m_X^2}\big(3s+8m_\psi^2\big) + 8\big(7s+32m_\psi^2\big)\bigg], \label{eq:A2_psiX}\\
\mathcal{A}^{(2)}_{A\to X} &= \frac{1}{15M_P^4}\bigg[\frac{3s^2}{m_X^4}\big(s^2+2m_A^2 s+6m_A^4\big) \nonumber\\
  &\quad + \frac{2s}{m_X^2}\big(13s^2+56m_A^2 s+48m_A^4\big)   \nonumber\\
  &\quad+ 4\big(37s^2+104m_A^2 s+192m_A^4\big)\bigg], \label{eq:A2_AX}\\
\mathcal{A}^{(2)}_{\gamma\to X} &= \frac{2s^2}{15M_P^4}\Big(\frac{s^2}{m_X^4}+\frac{12s}{m_X^2}+56\Big). \label{eq:A2_gamX}
\end{align}
The minimal terms follow the conventions of Ref.~\cite{Wang_2025}; see also Ref.~\cite{Tang_2017}
for related results. The terms of order $\xi$ and $\xi^2$ are new; for $m_i\to0$ they satisfy
$\mathcal{A}_{\gamma\to X}=4\,\mathcal{A}_{\psi\to X}$ order by order in $\xi$, and
$\mathcal{A}\geq0$ for all $\xi$. Note that $\mathcal{A}_{\gamma\to X}$ is not the $m_A\to0$ limit of
$\mathcal{A}_{A\to X}$: the longitudinal mode of $A_\mu$ does not decouple in this limit,
so that $\mathcal{A}_{A\to X}\to 29s^2/(15M_P^4)$ while
$\mathcal{A}_{\gamma\to X}\to 26s^2/(15M_P^4)$ for $m_X\to0$.

The corresponding cross section is
\begin{equation}
\sigma_{i\to X} = \frac{\mathcal{A}_{i\to X}}{64\pi s\, g_i^2}\,
  \sqrt{\frac{s-4m_X^2}{s-4m_i^2}} ,
\label{eq:sigma}
\end{equation}
where a symmetry factor $1/2$ for the identical final-state particles is included, and $g_i$ is the
number of internal degrees of freedom of each initial particle: $g_\varphi=1$, $g_\psi=2$,
$g_A=3$, and $g_\gamma=2$.

Above the electroweak crossover, $T_{\rm EW}\simeq 160$~GeV, the Higgs vacuum
expectation value vanishes and all SM particles are massless, up to thermal
corrections of order $gT$ which we neglect. We assume that the thermal bath
contains only SM states, namely four real scalars (the Higgs doublet), 45 Weyl
fermions (three generations without right-handed neutrinos), and 12 massless
gauge bosons. Weighting each contribution by its multiplicity $N_i$ and by the
initial-state symmetry factor $S_i$ defined in Eq.~\eqref{eq:rate_sigma} below,
we define
\begin{align}
\mathcal{A}_{{\rm SM}\to X} &\equiv \sum_i N_i S_i\,\mathcal{A}_{i\to X} \nonumber\\
  &= 2\mathcal{A}_{\varphi\to X}
  + \frac{45}{2}\mathcal{A}_{\psi\to X} + 6\mathcal{A}_{\gamma\to X} \nonumber\\
  &= \frac{3}{20M_P^4}\big(137\, s^2 + 584\, m_X^2 s + 512\, m_X^4\big) \nonumber\\
  &\quad - \frac{\xi}{30M_P^4}\bigg(\frac{101\, s^3}{m_X^2} + 4892\, s^2 + 1616\, m_X^2 s\bigg) \nonumber\\
  &\quad + \frac{\xi^2}{60M_P^4}\bigg(\frac{101\, s^4}{m_X^4} + \frac{1132\, s^3}{m_X^2} + 5496\, s^2\bigg) ,
\label{eq:A_SMX}
\end{align}
where $\mathcal{A}_{\varphi\to X}$, $\mathcal{A}_{\psi\to X}$, and
$\mathcal{A}_{\gamma \to X}$ are evaluated in the massless limit and denote the
contributions of a real scalar, a Dirac fermion (two Weyl fermions), and a massless
vector boson, respectively. Since the gravitational production rate scales as
$T^8/M_P^4$ (and even more steeply for $\xi\neq0$), as we show below, DM production is
dominated by the highest temperatures
of the thermal history and therefore takes place deep in the electroweak-symmetric
phase, provided the maximal bath temperature satisfies $T_{\max}\gg T_{\rm EW}$. For this
reason, the massive-vector contribution $\mathcal{A}_{A\to X}$ of
Eq.~\eqref{eq:A_AX} does not enter Eq.~\eqref{eq:A_SMX}. It would be required
only below $T_{\rm EW}$, where the $W^\pm$ and $Z$ bosons acquire masses by
absorbing three of the four Higgs degrees of freedom.

\subsection{Dark Matter Production}

Since the DM interacts only gravitationally, its abundance always remains
far below equilibrium, $n_X \ll n_X^{\rm eq}$, and inverse processes can be
neglected. The DM number density then obeys
\begin{equation}
    \frac{1}{a^3}\frac{\mathrm{d}}{\mathrm{d}t}\left(n_X a^3\right)
    = \mathcal{R}(T),
    \label{eq:boltzmann}
\end{equation}
where $\mathcal{R}(T)$ is the production rate per unit volume. For a
general initial species $i$ with mass $m_i$ and $g_i$ internal degrees of
freedom, approximating the initial-state distributions by
Maxwell--Boltzmann statistics gives~\cite{Gondolo:1990dk}
\begin{multline}
\mathcal{R}_i(T) = \frac{S_i\, g_i^2\, T}{16\,\pi^4}
  \int_{s_{\min}}^{\infty}\mathrm{d}s\;\sigma_{i\to X}(s) \\
  \times\left(s-4m_i^2\right)\sqrt{s}\;
  K_1\!\left(\frac{\sqrt{s}}{T}\right),
\label{eq:rate_sigma}
\end{multline}
where $s_{\min} = 4\max(m_i^2, m_X^2)$ and $K_1$ is the modified Bessel
function of the second kind. A factor of $2$ has been included because
each annihilation produces two DM particles, and $S_i$ is the symmetry
factor of the initial state: $S_i = 1/2$ for identical particles (real
scalars and vector bosons), and $S_i = 1$ for a Dirac fermion, since $\psi$
and $\bar\psi$ are distinct. For a bath containing $N_i$ species of type
$i$, the total rate is $\mathcal{R}(T) = \sum_i N_i\,\mathcal{R}_i(T)$.

It is convenient to express the result in terms of the yield
$Y_X \equiv n_X/\mathfrak{s}$, where
$\mathfrak{s} = \frac{2\pi^2}{45}\, g_{*s}\, T^3$ is the entropy density of
the bath. Introducing the number of e-folds $N_e$, defined by
$\mathrm{d}N_e \equiv H\,\mathrm{d}t$ so that $a \propto e^{N_e}$, and
assuming a vanishing DM abundance at the initial time $N_{e,0}$,
Eq.~\eqref{eq:boltzmann} integrates to
\begin{equation}
Y_X(N_e) = \frac{45}{2\pi^2 g_{*s}}\,\frac{e^{-3N_e}}{T^3(N_e)}
\int_{N_{e,0}}^{N_e} \mathcal{I}_X(N_e')\,\mathrm{d}N_e' ,
\label{eq:yield_integral}
\end{equation}
where
\begin{equation}
\mathcal{I}_X(N_e) \equiv e^{3N_e}\,\frac{\mathcal{R}\big(T(N_e)\big)}{H(N_e)}
\label{eq:IX}
\end{equation}
is the production rate of the comoving DM number density
$\mathcal{N}_X \equiv e^{3N_e} n_X$, i.e.\
$\mathcal{I}_X = \mathrm{d}\mathcal{N}_X/\mathrm{d}N_e$. Note that during
WI the entropy of the bath is not conserved, since radiation is
continuously sourced by the inflaton, so $Y_X$ only freezes to its final
value after WI ends.

The main difference between cold and warm inflation lies in when, and from what, dark matter can
be produced gravitationally. In cold inflation, the Universe supercools during inflation, and no
thermal bath is present~\cite{Berera:1995ie}. After inflation ends, the inflaton condensate
oscillates coherently about the minimum of its potential, and its quanta can be regarded as
particles at rest with energy $m_\phi$. During this reheating stage, DM is produced
gravitationally through condensate scattering, $\phi\phi\to XX$, which is kinematically allowed
only for $m_X<m_\phi$~\cite{Ema:2018ucl,Mambrini:2021zpp}. The SM bath, in contrast, does not
exist during inflation; it is created only during reheating, as the inflaton
decays~\cite{Allahverdi:2010xz}. The scattering $\mathrm{SM}\,\mathrm{SM}\to XX$ therefore
operates only from reheating onward, and it is typically subdominant, because the inflaton
dominates the energy density throughout this stage~\cite{Mambrini:2021zpp,Clery:2021bwz}.

\begin{figure}
  \includegraphics[width=0.8\columnwidth]{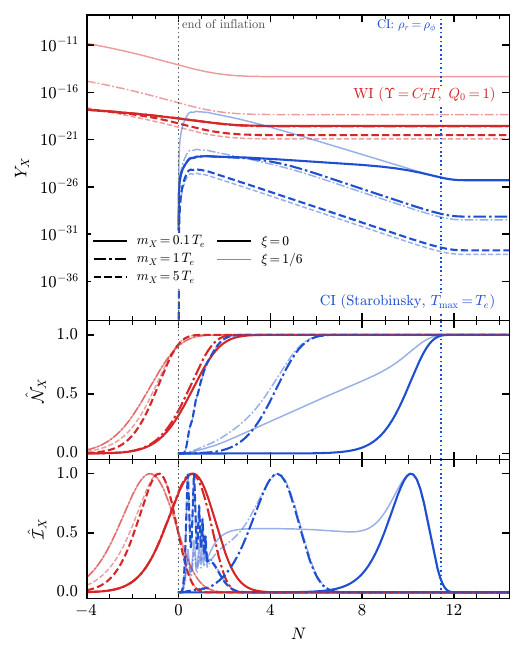}
  \caption{Gravitational freeze-in of vector DM from the SM bath for $\xi=0$ (thick) and
  $\xi=1/6$ (faint), in warm inflation with $Q_0=1$, $T_e=9.1\times10^{-6}M_P$ (red) and
  cold Starobinsky inflation with $T_{\max}=T_e$ (blue). Solid, dash-dotted and dashed lines
  correspond to $m_X=0.1,\,1,\,5\,T_e$. Top: yield $Y_X$. Middle: normalized comoving number
  $\hat{\mathcal{N}}_X$. Bottom: normalized production rate per e-fold $\hat{\mathcal{I}}_X$.
  Dotted lines mark the end of inflation ($N=0$) and $\rho_r=\rho_\phi$ in cold inflation.
  For $m_X=0.1\,T_e$ and $\xi=1/6$, $X$ has a transient gradient instability in cold
  inflation at $0.16\lesssim N\lesssim0.63$.}
  \label{fig:yield}
\end{figure}

In warm inflation, dissipation sustains a thermal bath from the beginning of inflation,
so DM is produced gravitationally through $\mathrm{SM}\,\mathrm{SM} \to XX$ throughout the
inflationary epoch, with production peaking around its end, as shown in Fig.~\ref{fig:yield}. The inflaton condensate, on
the other hand, never oscillates. For $\Upsilon \propto T$ it is overdamped after
inflation and relaxes smoothly towards its minimum, so the condensate scattering
$\phi\phi \to XX$ is absent. Inflaton quanta may be thermalized by the bath, in which case
they contribute as one additional bosonic species. For $T \gg m_{\rm eff}$, with
$m_{\rm eff}^2 \equiv V''(\phi)$, their contribution is at the percent level and can be
neglected.

\textbf{SM annihilation.---}Let us warm up by studying dark matter production through the
annihilation of SM particles. Inserting Eqs.~\eqref{eq:sigma} and \eqref{eq:A_SMX} into
Eq.~\eqref{eq:rate_sigma}, the factors $g_i^2$ cancel and the total rate becomes
\begin{multline}
    \mathcal{R}_{{\rm SM}\to X}(T) = \frac{T}{1024\,\pi^5}
    \int_{4m_X^2}^{\infty}\mathrm{d}s\;\mathcal{A}_{{\rm SM}\to X}(s)\, \\ \times
    \sqrt{s-4m_X^2}\;K_1\!\left(\frac{\sqrt{s}}{T}\right),
    \label{eq:R_SM}
\end{multline}
which we then use to evaluate the dark matter yield from Eq.~\eqref{eq:yield_integral}.
In the two limiting regimes the production rate is
\begin{align}
\mathcal{R}^{(m_X\ll T)}_{{\rm SM}\to X} &\simeq \frac{T^8}{\pi^5M_P^4}\bigg[\frac{1233}{80}
   -\xi\Big(\frac{606}{5}\frac{T^2}{m_X^2}+\frac{469}{4}\Big) \nonumber\\
 &\quad +\xi^2\Big(4848\,\frac{T^4}{m_X^4}+558\,\frac{T^2}{m_X^2}+\frac{303}{8}\Big)\bigg], \label{eq:R_SM_light}\\[2mm]
\mathcal{R}^{(m_X\gg T)}_{{\rm SM}\to X} &\simeq \frac{189-760\,\xi+776\,\xi^2}{128\pi^4}\,
   \frac{T^3m_X^5}{M_P^4}\,e^{-2m_X/T}. \label{eq:R_SM_heavy}
\end{align}
Here $T$ is the instantaneous temperature of the bath, $T=T(N_e)$, so that
$\mathcal{R}_{{\rm SM}\to X}$ is evaluated along the background in
Eq.~\eqref{eq:yield_integral}. 

Moreover, in the light-mass limit $m_X\ll T$ the rate in Eq.~\eqref{eq:R_SM_light} is a sum
of terms of the form $\mathcal{R}=T^{2n+4}/\Lambda^{2n}$, characteristic of UV freeze-in through
non-renormalizable interactions~\cite{Freese_2024}. Keeping the leading term at each order
in $\xi$,
\begin{align}
\mathcal{R}_{{\rm SM}\to X} &\simeq \frac{T^8}{\Lambda_2^4}
  \mp\frac{T^{10}}{\Lambda_3^6}+\frac{T^{12}}{\Lambda_4^8}, \label{eq:R_UV}\\
\Lambda_2 &= \Big(\frac{80\pi^5}{1233}\Big)^{1/4}M_P\simeq 2.1\,M_P, \nonumber\\
\Lambda_3 &= \Big(\frac{5\pi^5}{606\,|\xi|}\,M_P^4m_X^2\Big)^{1/6}
  \simeq 1.2\,\frac{(M_P^2m_X)^{1/3}}{|\xi|^{1/6}}, \nonumber\\
\Lambda_4 &= \Big(\frac{\pi^5}{4848\,\xi^2}\,M_P^4m_X^4\Big)^{1/8}
  \simeq 0.71\,\frac{(M_Pm_X)^{1/2}}{|\xi|^{1/4}}, \nonumber
\end{align}
where the upper (lower) sign applies to $\xi>0$ ($\xi<0$).
The minimal-coupling term ($n=2$) shows that graviton exchange acts as a Planck-suppressed
operator of mass dimension $n+4=6$. The non-minimal coupling adds effective operators of
dimension $7$ and $8$, whose scales $\Lambda_{3,4}$ decrease as $m_X$ decreases. In both cases
production is dominated by the highest temperature reached. The EFT remains valid as long as the
typical collision energy, $\sqrt{s}\simeq10\,T$ for the $\xi^2$ term, stays below $\Lambda_4$, which
is also the unitarity scale. Including all SM channels, this requires
$|\xi|\lesssim3.5\times10^{-6}\,M_P^2m_X^2/T_{\max}^4$.

\textbf{Inflaton annihilation.---} We decompose the inflaton into its homogeneous condensate and
its fluctuations, $\phi(t,\mathbf x)=\bar\phi(t)+\delta\phi(t,\mathbf x)$. In cold inflation the
fluctuations $\delta\phi$ remain in their vacuum state; stretched beyond the horizon, they seed the
curvature perturbations and hence all cosmic structure~\cite{Mukhanov:1990me}. After inflation, the condensate oscillates
coherently about the minimum of its potential and produces dark matter through $\bar\phi\bar\phi\to h^*\to XX$, which is kinematically allowed for $m_X<m_\phi$ and typically
dominates over production from the SM bath during
reheating~\cite{Garcia:2023qab}. Evaluating $\mathcal{A}_{\varphi\to X}$ at $s=4m_\phi^2$, its
production rate is
\begin{align}
\mathcal{R}^{\rm CI}_{\phi\to X} &= \frac{\rho_\phi^2\sqrt{1-x^2}}{128\pi M_P^4}
  \bigg[1-x^2+\frac34x^4 \nonumber\\
 &\quad -4\xi\Big(\frac{2}{x^2}-1+2x^2\Big)
  +16\xi^2\Big(2+\frac{1}{x^4}\Big)\bigg],
\label{eq:R_phi_X}
\end{align}
Here $x\equiv m_X/m_\phi$ and $\rho_\phi$ is the oscillation-averaged energy density of the condensate.

In warm inflation the picture changes in two ways. First, the coherent channel is absent: strong
dissipation overdamps the condensate, so in our benchmark models it relaxes to the minimum without
oscillating. Second, if the inflaton fluctuations are thermalized, as assumed in computing
$\Delta_\mathcal{R}^2$ in Fig.~\ref{fig:cmb_constraints}, the quanta $\delta\phi$ belong to the
thermal bath and produce dark matter like any SM species, through $\delta\phi\,\delta\phi\to h^*\to XX$. Assuming $m_\phi \ll T$, as a single real scalar, they contribute
\begin{align}
\mathcal{R}^{\mathrm{WI},(m_X\ll T)}_{\delta\phi\to X} &\simeq \frac{T^8}{\pi^5M_P^4}\bigg[\frac{3}{40}
   -\xi\Big(\frac{12}{5}\frac{T^2}{m_X^2}+\frac{1}{4}\Big) \nonumber\\
 &\quad +\xi^2\Big(96\,\frac{T^4}{m_X^4}+\frac{3}{4}\Big)\bigg], \label{eq:R_dphi_light}\\[2mm]
\mathcal{R}^{\mathrm{WI},(m_X\gg T)}_{\delta\phi\to X} &\simeq \frac{3-16\,\xi+32\,\xi^2}{512\pi^4}\,
   \frac{T^3m_X^5}{M_P^4}\,e^{-2m_X/T}.
   \label{eq:R_dphi_heavy}
\end{align}
For minimal coupling these are about $0.5\%$ and $0.4\%$ of $\mathcal{R}_{\rm SM\to X}$,
respectively, so in warm inflation dark matter is produced almost entirely by the SM bath.

Figure~\ref{fig:yield} shows that, for both $\xi=0$ and $\xi=1/6$, freeze-in from the SM bath
is far more efficient in warm inflation than in cold inflation. This is because in warm
inflation a hot radiation bath is sustained during inflation itself, so graviton-mediated
scattering, whose rate grows steeply with temperature ($\mathcal{R}\propto T^{8}$ for
$m_X\ll T$, and up to $T^{12}$ when the $\xi^2$ term dominates), operates at $T\gtrsim T_e$ for
several e-folds. In cold inflation, by contrast,
the bath appears only after inflation and stays near $T_{\max}$ only briefly during reheating.
We stress, however, that this comparison includes only production from the SM bath,
${\rm SM}\,{\rm SM}\to h^*\to XX$. Gravitational scattering of the inflaton condensate,
$\phi\phi\to h^*\to XX$, is not included, although it typically dominates gravitational DM
production during reheating in cold inflation.

\begin{figure}[t]
    \centering
    \includegraphics[width=0.8\columnwidth]{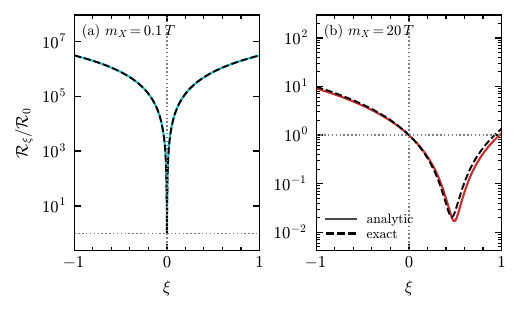}
    \caption{Ratio of the freeze-in rate from the SM bath with non-minimal coupling $\xi$ to
    the minimal one, $\mathcal{R}_\xi/\mathcal{R}_0$, as a function of $\xi$.
    (a)~Light dark matter, $m_X=0.1\,T$; (b)~heavy dark matter, $m_X=20\,T$.
    Solid lines show the analytic limits of Eq.~\eqref{eq:ratio} and dashed lines the exact
    thermal integral. For light $X$ the longitudinal mode enhances the rate as
    $\propto\xi^2T^4/m_X^4$, while for heavy $X$ destructive interference between graviton
    exchange and the contact term suppresses it near $\xi\simeq0.5$.}
    \label{fig:ratio_xi}
\end{figure}

Next, we discuss how $\xi$ modifies the collision term in the Boltzmann equation. The ratio
of the rates with and without $\xi$, shown in Fig.~\ref{fig:ratio_xi}, is
\begin{subequations}\label{eq:ratio}
\begin{align}
\frac{\mathcal{R}_\xi}{\mathcal{R}_0} &\simeq 1-\xi\Big(7.86\,\frac{T^2}{m_X^2}+7.61\Big) \nonumber\\
 &\quad +\xi^2\Big(315\,\frac{T^4}{m_X^4}+36.2\,\frac{T^2}{m_X^2}+2.46\Big), \label{eq:ratio_light}\\
\frac{\mathcal{R}_\xi}{\mathcal{R}_0} &\simeq 1-4.02\,\xi+4.11\,\xi^2, \label{eq:ratio_heavy}
\end{align}
\end{subequations}
for $m_X\ll T$ and $m_X\gg T$, respectively, where
$\mathcal{R}_\xi\equiv\mathcal{R}_{{\rm SM}\to X}(\xi)$.
For light $X$ the minimum lies essentially at $\xi\simeq0$, and the rate grows as
$\xi^2T^4/m_X^4$ because the contact term couples to the longitudinal mode,
$\epsilon_L\sim\sqrt{s}/m_X$. For heavy $X$, production occurs near threshold, where graviton
exchange and the contact term in $\mathcal{M}=\mathcal{M}_h+\xi\,\mathcal{M}_c$ are
comparable and interfere destructively, ${\rm Re}(\mathcal{M}_h^*\mathcal{M}_c)<0$, giving a
minimum $\mathcal{R}_\xi/\mathcal{R}_0\simeq0.02$ at $\xi\simeq0.5$.

\subsection{Relic abundance from freeze-in}
\label{sec:relic_freezein}

Having established how the yield of $X$ evolves, we now require that $X$ accounts for all of the
dark matter, $\Omega_Xh^2=0.12$~\cite{Planck:2018vyg}. Once production has stopped, the yield
$Y_X^\infty$ is conserved, and the present abundance is
\begin{equation}
\Omega_Xh^2=\frac{m_X\,Y_X^\infty\,s_0\,h^2}{\rho_c}
\simeq2.74\times10^{8}\left(\frac{m_X}{\rm GeV}\right)Y_X^\infty ,
\label{eq:omega}
\end{equation}
where $s_0=\frac{2\pi^2}{45}g_{*s}(t_0)T_0^3$ with $g_{*s}(t_0)=43/11$ and $T_0=2.725$~K is the
present entropy density, and $\rho_c/h^2=1.05\times10^{-5}~{\rm GeV\,cm^{-3}}$.

We first derive $Y_X^\infty$ analytically in the light ($m_X\ll T_e$) and heavy ($m_X\gg T_e$)
limits, and then compare with the full numerical solution.
Approximating the background by $T\simeq T_e$ for $a\le a_e$, $T\simeq T_e\,a_e/a$ afterwards, and
$\rho\simeq2\rho_r$ at the end of warm inflation (since $\rho_\phi\simeq\rho_r$), so that
$H=\pi\sqrt{g_*/45}\,T^2/M_P$, Eq.~\eqref{eq:yield_integral} with the rates of
Eq.~\eqref{eq:ratio} gives
\begin{subequations}
\label{eq:Y_analytic}
\begin{align}
Y_X^\infty&\simeq1.5\times10^{-4}\,\mathcal{F}_\xi^{\rm L}(x_e)\left(\frac{T_e}{M_P}\right)^{3},
\label{eq:Y_light}\\
Y_X^\infty&\simeq2.2\times10^{-5}\,\mathcal{F}_\xi^{\rm H}\,x_e^{5}\,e^{-2x_e} \nonumber\\
&\quad\times\Big(1+\frac{3}{2x_e}\Big)\left(\frac{T_e}{M_P}\right)^{3},
\label{eq:Y_heavy}
\end{align}
\end{subequations}
for $m_X\ll T_e$ and $m_X\gg T_e$, respectively, with $x_e=m_X/T_e$, $g_*=106.75$, and
\begin{subequations}
\label{eq:F_xi}
\begin{align}
\mathcal{F}_\xi^{\rm L} &=1-\xi\Big(\frac{6.29}{x_e^2}+7.61\Big) \nonumber\\
&\quad+\xi^2\Big(\frac{225}{x_e^4}+\frac{29.0}{x_e^2}+2.46\Big),
\label{eq:F_light}\\
\mathcal{F}_\xi^{\rm H}&=1-4.02\,\xi+4.11\,\xi^2 .
\label{eq:F_heavy}
\end{align}
\end{subequations}
In Eq.~\eqref{eq:Y_heavy} the first term is produced during inflation and the second after
it. In Eq.~\eqref{eq:Y_light} the two epochs contribute equally for $\xi=0$, while the
$\xi$-enhanced terms, $\propto T^{10}$ and $T^{12}$, are produced mostly during inflation (5/8
and 7/10 of the total, respectively), since they are more strongly UV dominated. Inserting
Eq.~\eqref{eq:Y_analytic} into Eq.~\eqref{eq:omega}, the relic abundance is
\begin{subequations}
\label{eq:omega_analytic}
\begin{align}
\frac{\Omega_Xh^2}{0.12}&\simeq8.4\times10^{23}\,\mathcal{F}_\xi^{\rm L}(x_e)\,
\frac{m_X\,T_e^{3}}{M_P^{4}},
\label{eq:omega_light}\\
\frac{\Omega_Xh^2}{0.12}&\simeq1.2\times10^{23}\,\mathcal{F}_\xi^{\rm H}\,x_e^{6}\,e^{-2x_e} \nonumber\\
&\quad\times\Big(1+\frac{3}{2x_e}\Big)\left(\frac{T_e}{M_P}\right)^{4}.
\label{eq:omega_heavy}
\end{align}
\end{subequations}
for light and heavy $X$, respectively. For $\xi\neq0$ the light case is extremely
sensitive to the non-minimal coupling: since $x_e\ll1$, the terms $\propto\xi/x_e^2$ and
$\xi^2/x_e^4$ in $\mathcal{F}^{\rm L}_\xi$ dominate as soon as
$|\xi|\gtrsim x_e^2/15$, after which $\Omega_Xh^2\propto\xi^2/m_X^3$ grows rapidly toward
smaller masses. Because the light solutions of Eq.~\eqref{eq:omega_light} lie at
$x_e\sim10^{-8}$--$10^{-3}$, even a tiny coupling removes them, e.g., $\xi=10^{-12}$
already eliminates the light branch for $Q_0\le10^{-1}$. In the heavy case, by contrast,
$\mathcal{F}^{\rm H}_\xi$ is independent of $x_e$ and of order unity for $|\xi|\lesssim1$,
so the heavy solution $m_X\simeq13$--$26\,T_e$ is essentially unaffected.

\begin{figure}
    \centering
    \includegraphics[width=0.8\columnwidth]{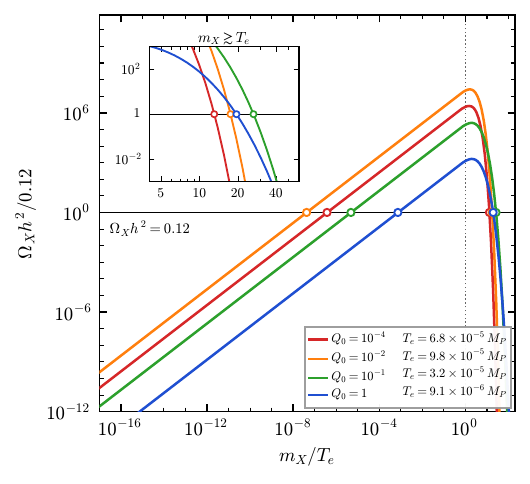}
    \caption{Relic abundance of $X$ from gravitational freeze-in, $\Omega_Xh^2/0.12$, as a function of
    $m_X/T_e$ for the CMB-allowed warm-inflation models, $\xi=0$. Colored curves show the benchmarks
    $Q_0=10^{-4},10^{-2},10^{-1},1$, and the shaded band spans all models with
    $10^{-4}\leq Q_0\leq1$. Open circles mark $\Omega_Xh^2=0.12$; the inset enlarges the heavy-mass
    branch, $m_X\gtrsim T_e$.}
    \label{fig:relic}
\end{figure}

\textbf{Minimal case, $\xi=0$.---} Moving to the numerical results, shown in Fig.~\ref{fig:relic}, the relic abundance of all
CMB-allowed models is well described by
\begin{multline}
\frac{\Omega_Xh^2}{0.12}\simeq(2.7\pm1.0)\times10^{23}\,  \\ \times\frac{m_X\,T_e^{3}}{M_P^{4}}
\exp\!\left[-c_1\left(\frac{m_X}{T_e}\right)^{c_2}\right],
\label{eq:omega_fit}
\end{multline}
with $c_1\simeq0.6$ and $c_2$ decreasing from $1.3$ ($Q_0=10^{-4}$) to $0.94$ ($Q_0=1$). It
covers both branches: for $m_X\ll T_e$ the exponential is unity and $\Omega_Xh^2\propto m_X$,
while for $m_X\gtrsim T_e$ it describes the Boltzmann suppression, reproducing the heavy relic
masses to within $4\%$ for $m_X\lesssim40\,T_e$.

Requiring $\Omega_Xh^2\leq0.12$ then constrains the DM mass to
\begin{subequations}
\label{eq:mass_bounds}
\begin{align}
m_X&\lesssim  9\times10^{-6}\,{\rm GeV}\left(\frac{M_P}{T_e}\right)^{3},
\label{eq:mass_light}\\
\text{or}\quad m_X&\gtrsim T_e\left[\frac{4}{c_1}\,\ln\!\left(1.5\times10^{6}\,\frac{T_e}{M_P}\right)\right]^{1/c_2},
\label{eq:mass_heavy}
\end{align}
\end{subequations}
where the equalities correspond to $X$ making up all of the dark matter, and we give a summary of mass range in Tab.~\ref{tab:mass_bounds}.

\begin{table}[t]
\caption{Dark matter mass bounds for minimal coupling, $\xi=0$, from $\Omega_Xh^2\leq0.12$ for the four warm-inflation benchmarks,
obtained from the full numerical solution. $T_e$ and $H_e$ are the temperature and Hubble rate at
the end of inflation. Dark matter is allowed for $m_X\leq m_X^{\rm light}$ or
$m_X\geq m_X^{\rm heavy}$; the equalities correspond to $X$ making up all of the dark matter.}
\label{tab:mass_bounds}
\renewcommand{\arraystretch}{1.35}
\begin{ruledtabular}
\begin{tabular}{ccccc}
$Q_0$ & $T_e$ & $H_e$ & $m_X^{\rm light}$ & $m_X^{\rm heavy}$\\[-3pt]
 & [$10^{-5}M_P$] & [$10^{-9}M_P$] & [$10^{7}$\,GeV] & [$10^{15}$\,GeV]\\
\hline
$10^{-4}$ & $6.83$  & $264$   & $6.2$  & $2.2$\\
$10^{-2}$ & $9.83$  & $62.6$  & $1.0$  & $4.2$\\
$10^{-1}$ & $3.18$  & $4.98$  & $37$   & $2.0$\\
$1$       & $0.906$ & $0.398$ & $1600$ & $0.43$\\
\end{tabular}
\end{ruledtabular}
\end{table}

\begin{figure}
    \centering
    \includegraphics[width=0.8\columnwidth]{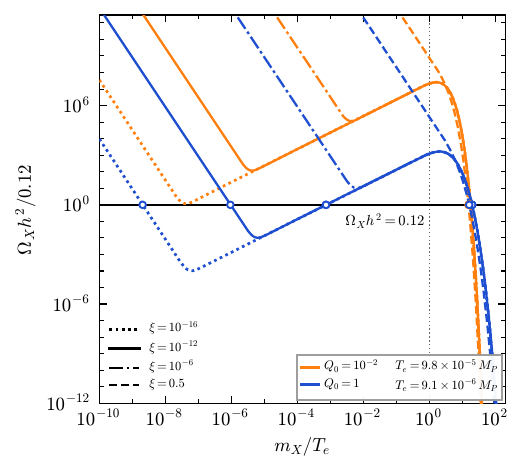}
    \caption{Relic abundance of vector DM from freeze-in from the SM bath in warm inflation
    with non-minimal coupling $\xi$, for $Q_0=10^{-2}$ (orange, $T_e=9.8\times10^{-5}M_P$) and
    $Q_0=1$ (blue, $T_e=9.1\times10^{-6}M_P$). Dotted, solid, dash-dotted and dashed lines
    correspond to $\xi=10^{-16},\,10^{-12},\,10^{-6}$ and $0.5$. Open circles mark
    $\Omega_Xh^2=0.12$, and the vertical dotted line marks $m_X=T_e$.}
    \label{fig:relic_xi}
\end{figure}

\textbf{Non-minimal case, $\xi\neq0$.---} In this case the coupling $\xi$ strongly affects the
light branch, as shown in Fig.~\ref{fig:relic_xi} for $Q_0=10^{-2}$ and $1$. For $\xi\neq0$
the abundance becomes V-shaped: toward smaller masses the $\xi^2$ term dominates and
$\Omega_Xh^2\propto\xi^2/m_X^3$, with the minimum at $m_X\sim\sqrt{\xi}\,T_e$. The light branch
therefore exists only for extremely weak couplings, $|\xi|\lesssim7\times10^{-17}$ for
$Q_0=10^{-2}$ and $|\xi|\lesssim10^{-8}$ for $Q_0=1$. Within this range it splits into two solutions, e.g.,
$m_X\simeq2\times10^{-9}\,T_e$ and $7\times10^{-4}\,T_e$ for $Q_0=1$ and $\xi=10^{-16}$.
The heavy branch, by contrast, exists for any $\xi$ and is nearly insensitive to it: it is
unchanged for $\xi\le10^{-6}$ and moves to $\sim\!15\%$ lower masses for $\xi=0.5$, owing to
the destructive interference near threshold. The numerical relic abundance of all CMB-allowed
models is well described by
\begin{multline}
\frac{\Omega_Xh^2}{0.12}\simeq(2.7\pm1.0)\times10^{23}\,\frac{m_X\,T_e^{3}}{M_P^{4}}
\exp\!\left[-c_1\Big(\frac{m_X}{T_e}\Big)^{c_2}\right]\\
\times\left[(1-2.2\,\xi)^2+\kappa\,\xi^2\,\frac{T_e^2\,(T_e^2+0.32\,m_X^2)}{m_X^4}\right],
\label{eq:omega_fit_xi}
\end{multline}
with $\kappa\simeq220$ ($80$--$570$ for $Q_0=10^{-4}$--$1$), and $c_1$, $c_2$ as in
Eq.~\eqref{eq:omega_fit}. At $\xi=0$ the bracket equals unity and
Eq.~\eqref{eq:omega_fit} is recovered.

For $|\xi|\le\xi^*$, the requirement $\Omega_Xh^2\leq0.12$ then gives
\begin{subequations}
\label{eq:mass_bounds_xi}
\begin{align}
m_X&\lesssim 9\times10^{-6}\,{\rm GeV}\left(\frac{M_P}{T_e}\right)^{3}, \nonumber\\
m_X&\gtrsim 9.5\times10^{26}\,{\rm GeV}\,\xi^{2/3}\left(\frac{T_e}{M_P}\right)^{7/3},
\label{eq:mass_light_xi}\\
\text{or}\quad m_X&\gtrsim T_e\left[\frac{4}{c_1}\,
\ln\!\left(1.5\times10^{6}\,\mathcal{G}_{\rm H}^{1/4}\,\frac{T_e}{M_P}\right)\right]^{1/c_2},
\label{eq:mass_heavy_xi}
\end{align}
\end{subequations}
where the lower light bound scales as $(\kappa/220)^{1/3}$ and
$\mathcal{G}_{\rm H}\simeq(1-2.2\,\xi)^2+0.32\,\kappa\,\xi^2T_e^2/m_X^2$. The two light roots
merge at
\begin{equation}
\xi^*\simeq3\times10^{-49}\Big(\frac{220}{\kappa}\Big)^{1/2}\left(\frac{M_P}{T_e}\right)^{8},
\label{eq:xi_star}
\end{equation}
and for $|\xi|>\xi^*$ only the heavy branch, Eq.~\eqref{eq:mass_heavy_xi}, survives.

We now determine the region of the $(m_X,\xi)$ plane that is consistent with the observed
dark-matter abundance. Each value of $Q_0$ gives a different warm-inflation background, with
different $H_e$ and $T_e$, while all models with $10^{-4}\le Q_0\le1$ are compatible with the
CMB constraints as shown in Fig.~\ref{fig:cmb_constraints}. Since the CMB does not single out a particular $Q_0$,
we compute the relic abundance on each of these backgrounds and take the intersection of the
regions with $\Omega_Xh^2\le0.12$. The result is a conservative allowed region, valid
independently of $Q_0$ within the CMB-allowed range. 

\begin{figure}
    \centering
    \includegraphics[width=0.8\columnwidth]{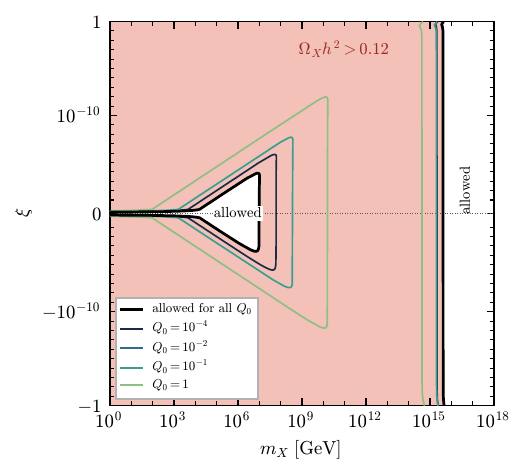}
    \caption{Allowed region in the $(m_X,\xi)$ plane for vector DM produced by gravitational
    freeze-in from the SM bath in warm inflation. Thin colored lines mark
    $\Omega_Xh^2=0.12$ for the CMB-compatible models $Q_0=10^{-4}$, $10^{-2}$, $10^{-1}$ and
    $1$. The red region is excluded by overproduction for at least one $Q_0$, and the thick
    black line bounds the region allowed for all $Q_0$ (white). The $\xi$ axis is logarithmic
    for $|\xi|>10^{-20}$ and linear below.}
    \label{fig:scan_mX_xi}
\end{figure}

As shown in Fig.~\ref{fig:scan_mX_xi}, the allowed mass also depends on the background $Q_0$.
In the minimal case, the light branch requires $m_X\lesssim1.6\times10^{10}$~GeV
($6.2\times10^{7}$~GeV) and the heavy branch $m_X\gtrsim4.3\times10^{14}$~GeV
($2.2\times10^{15}$~GeV) for $Q_0=1$ ($10^{-4}$). In the non-minimal case, the light branch
survives only for $|\xi|\lesssim10^{-8}$ ($2\times10^{-15}$), while the heavy branch is nearly
unchanged, with $m_X\gtrsim3.9\times10^{15}$~GeV for all $Q_0$.

\section{Cosmological Production}
\label{sec:CGPP}

We now consider DM production through vacuum excitation in the expanding
Universe, which arises from the zeroth-order term in $1/M_P$ of
Eq.~\eqref{eq:action_expanded}, now including the non-minimal coupling
$\xi G_{\mu\nu}X^\mu X^\nu$ evaluated on the background.

In conformal time, $g_{\mu\nu} = a^2(\eta)\,\eta_{\mu\nu}$, where
$\eta_{\mu\nu}$ is the Minkowski metric and $\mathrm{d}t = a\,\mathrm{d}\eta$.
On this background the Einstein tensor is $G_{00}=3\mathcal{H}^2$ and
$G_{ij}=-(2\mathcal{H}'+\mathcal{H}^2)\,\delta_{ij}$, with $\mathcal{H}\equiv a'/a=aH$.
Since the Maxwell term is conformally invariant and $G_{\mu\nu}$ contains no
derivatives of $X_\mu$, the action becomes
\begin{multline}
    S_X = \int \mathrm{d}^4x \bigg[
        -\frac{1}{4}X^{\mu\nu}X_{\mu\nu}
        + \frac{1}{2}M_t^2\,X_0^2 \\
        - \frac{1}{2}M_s^2\,\delta^{ij}X_iX_j \bigg],
    \label{eq:action_conformal}
\end{multline}
with indices raised by $\eta^{\mu\nu}$, where the temporal and spatial
components acquire different effective masses,
\begin{subequations}\label{eq:Mts}
\begin{align}
    M_t^2 &\equiv a^2m_X^2 + 6\xi\mathcal{H}^2 = a^2\big(m_X^2+6\xi H^2\big), \\
    M_s^2 &\equiv a^2m_X^2 + 2\xi\big(2\mathcal{H}'+\mathcal{H}^2\big) \nonumber\\
          &= a^2\big(m_X^2-6w\,\xi H^2\big),
\end{align}
\end{subequations}
and $w=p/\rho$ is the background equation of state. Conformal symmetry is
broken by the mass term and, for $\xi\neq0$, by the curvature. During
inflation, $w\simeq-1$ and $G_{\mu\nu}\simeq-3H^2g_{\mu\nu}$, so that
$M_t^2=M_s^2=a^2(m_X^2+6\xi H^2)$ and the coupling acts as a common mass shift.
After inflation, $w\neq-1$ and the two masses differ. For $\xi=0$ both reduce
to $a^2m_X^2$.

Owing to the spatial homogeneity of the background, we expand $X_\mu$ in
Fourier modes,
\begin{equation}
    X_{\mu}(\eta, \mathbf{x}) = \int \frac{\mathrm{d}^3k}{(2\pi)^{3/2}}\,
    \tilde{X}_{\mu}(\eta,\mathbf{k})\, e^{i \mathbf{k} \cdot \mathbf{x}},
    \label{eq:fourier}
\end{equation}
with $\tilde{X}_\mu(\eta,-\mathbf{k}) = \tilde{X}^*_\mu(\eta,\mathbf{k})$.
The spatial part is decomposed into transverse and longitudinal components,
\begin{equation}
    \tilde{\mathbf{X}} = \tilde{\mathbf{X}}_T + \tilde{X}_L\,\hat{\mathbf{k}},
    \qquad \mathbf{k}\cdot\tilde{\mathbf{X}}_T = 0 .
\end{equation}
The non-dynamical $\tilde{X}_0$ is fixed by the constraint
\begin{equation}
    \tilde{X}_0 = -\frac{ik\,\tilde{X}'_L}{k^2 + M_t^2}.
    \label{eq:constraint}
\end{equation}
Substituting Eq.~\eqref{eq:constraint}, the action splits into decoupled
transverse and longitudinal parts,
\begin{multline}
    S = \frac{1}{2}\int \mathrm{d}\eta\, \mathrm{d}^3k \,\bigg[
        |\tilde{\mathbf{X}}'_T|^2 - \left(k^2 + M_s^2\right)|\tilde{\mathbf{X}}_T|^2 \\
        + \frac{M_t^2}{k^2 + M_t^2}\,|\tilde{X}'_L|^2
        - M_s^2|\tilde{X}_L|^2 \bigg].
    \label{eq:action_modes}
\end{multline}

\textbf{Transverse modes.---} We expand
\begin{multline}
    \tilde{\mathbf{X}}_T(\eta,\mathbf{k}) =  \sum_{\lambda=\pm} \Big[
        \hat{a}_{\lambda}(\mathbf{k})\, v_T(\eta,k)\,
        \boldsymbol{\epsilon}_\lambda(\hat{\mathbf{k}}) \\
      + \hat{a}^\dagger_{\lambda}(-\mathbf{k})\, v^*_T(\eta,k)\,
        \boldsymbol{\epsilon}^*_\lambda(-\hat{\mathbf{k}}) \Big],
    \label{eq:XT_expansion}
\end{multline}
where the mode function obeys
\begin{equation}
    v_T'' + \omega_T^2\, v_T = 0, \qquad
    \omega_T^2 = k^2 + M_s^2 ,
    \label{eq:vT_eom}
\end{equation}
with the Wronskian normalization $v_T v_T'^* - v_T' v_T^* = i$.

\textbf{Longitudinal modes.---} The longitudinal kinetic term in
Eq.~\eqref{eq:action_modes} is not canonically normalized. We therefore
introduce
\begin{equation}
    \mathcal{X}_L = f\,\tilde{X}_L, \qquad
    f(\eta,k) \equiv \frac{M_t}{\sqrt{k^2 + M_t^2}} ,
    \label{eq:f_def}
\end{equation}
in terms of which, after integrating by parts, the action takes the
canonical form. Expanding
\begin{equation}
    \mathcal{X}_L = \hat{a}_L(\mathbf{k})\, v_L(\eta,k) + \hat{a}^\dagger_L(-\mathbf{k})\, v^*_L(\eta,k),
\end{equation}
the mode function obeys $v_L'' + \omega_L^2\, v_L = 0$ with
\begin{equation}
    \omega_L^2 = \frac{M_s^2}{M_t^2}\left(k^2 + M_t^2\right) - \frac{f''}{f},
    \label{eq:vL_eom}
\end{equation}
and the same Wronskian normalization. For $\xi=0$ this reduces to
$\omega_L^2=k^2+a^2m_X^2-f''/f$, and the longitudinal mode differs from the
transverse ones only through $f''/f$. In the relativistic regime,
$k \gg M_t$, one has $f \simeq M_t/k$, and the longitudinal mode propagates with
sound speed $c_L^2=M_s^2/M_t^2$. In the non-relativistic regime,
$k \ll M_t$, $f \to 1$ and $f''/f$ is suppressed. For $\xi=0$ and $m_X \to 0$
the transverse modes become conformal and are not produced, while the
longitudinal mode does not decouple.

\textbf{Stability.---} The longitudinal mode is free of ghosts only if $M_t^2>0$, and free of a
gradient instability only if $c_L^2=M_s^2/M_t^2>0$. If either condition is violated, high-$k$
modes grow without bound, leading to runaway particle production~\cite{Capanelli:2024pzd,
Capanelli:2024rlk}. Using Eq.~\eqref{eq:Mts}, the two conditions read $m_X^2+6\xi H^2>0$ and
$m_X^2-6w\,\xi H^2>0$. For $\xi<0$ the first condition is strongest during inflation, where $H$ is
largest. For $\xi>0$ only the second condition matters, and only in epochs with $w>0$, such as
radiation domination or the oscillation phase of cold inflation. In warm inflation $w<0$ during
inflation and $w\leq1/3$ afterwards, so the two conditions together give
\begin{equation}
-\frac{m_X^2}{6H_*^2}\lesssim\xi\lesssim\frac{m_X^2}{2H_e^2},
\label{eq:stability}
\end{equation}
where $H_*$ is the Hubble rate at horizon exit of the CMB pivot scale. The upper bound is weaker
than in cold inflation, where the instantaneous equation of state reaches $w\simeq1$ whenever the
oscillating inflaton passes through the minimum of its potential, giving
$\xi\lesssim m_X^2/(6H^2)$. In addition, $c_L>1$ for $\xi<0$ whenever $w>-1$, so the longitudinal
mode then propagates superluminally. On the heavy branch, $m_X\gg T_e\gg H_e$, so the upper bound
is satisfied for all couplings of interest in Sec.~\ref{sec:relic_freezein}. For $\xi<0$, however,
the ghost condition gives $|\xi|\lesssim3.6\,x_e^2$ for $Q_0=1$, with $x_e\equiv m_X/T_e$, and must
be imposed in addition.

\textbf{Particle number.---} We impose Bunch--Davies initial conditions deep
inside the horizon, $k \gg aH$. There $M_t^2=M_s^2$ (since $w\simeq-1$),
$\omega_{T,L} \to k$, and the evolution is adiabatic,
\begin{equation}
    v_k(\eta_i) = \frac{1}{\sqrt{2\omega_k}}, \quad
    v_k'(\eta_i) = -\left(i\omega_k + \frac{\omega_k'}{2\omega_k}\right)
    v_k(\eta_i).
\end{equation}
We evolve Eqs.~\eqref{eq:vT_eom} and \eqref{eq:vL_eom} numerically into the
radiation-dominated era, until $|\omega_k'/\omega_k^2| \ll 1$ and
$|\xi|H^2\ll m_X^2$, so that the curvature terms have switched off. There we
evaluate the particle number
\begin{equation}
    n_k = |\beta_k|^2
    = \frac{|v_k'|^2 + \omega_k^2 |v_k|^2}{2\omega_k} - \frac{1}{2},
    \label{eq:nk}
\end{equation}
and the comoving DM number density
\begin{equation}
    a^3 n_X = \int_0^{\infty} \mathrm{d} \ln{k}~\frac{k^3}{2\pi^2}
    \left(n_k^{(L)} + 2\,n_k^{(T)}\right) .
    \label{eq:nX_CGPP}
\end{equation}

\subsection{Comoving number spectrum}
\label{sec:cgpp_spectrum}

\begin{figure}
    \centering
    \includegraphics[width=0.9\linewidth]{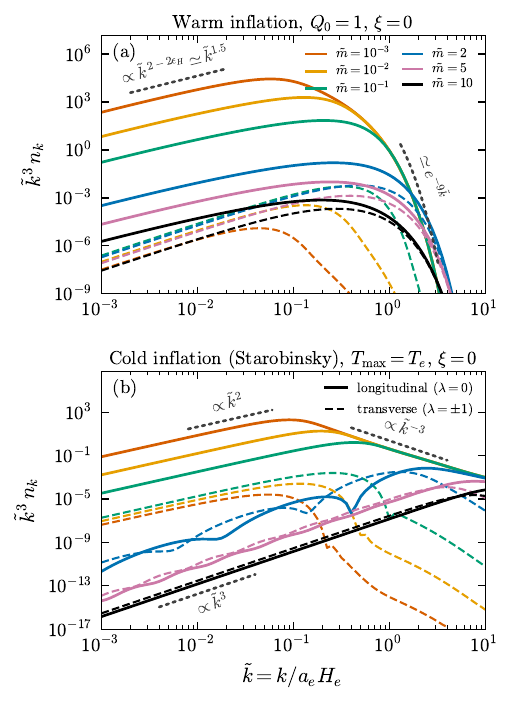}
    \caption{Gravitational production spectra $\tilde k^3 n_k$ of a minimally coupled ($\xi=0$) Proca field for (a) warm inflation with $Q_0=1$ and (b) Starobinsky cold inflation with $T_{\max}=T_e$, in units $a_e=H_e=1$. Colors label $\tilde m=m_X/H_e$; solid (dashed) lines show the longitudinal (transverse) modes. Dotted lines indicate the asymptotic slopes. The longitudinal spectrum is suppressed at low $k$, and in warm inflation it is cut off exponentially at high $k$.}
    \label{fig:cgpp_spectrum}
\end{figure}

In this subsection, we evaluate Eq.~\eqref{eq:nk} numerically at late times, $\eta\to+\infty$, for several dark matter
masses. Figure~\ref{fig:cgpp_spectrum} shows the resulting spectra $\tilde k^3 n_k$. 

In warm inflation, Fig.~\ref{fig:cgpp_spectrum}(a), the longitudinal spectrum of light dark matter ($m_X\ll H_e$) grows on large
scales as
\begin{equation}
\tilde k^3 n_k\propto\tilde k^{\,2-2\epsilon_H},
\quad
\epsilon_H\equiv-\frac{\dot H}{H^2}=\frac{\dot\phi^2}{2M_P^2H^2}+\frac{2\rho_r}{\rho},
\label{eq:IR_slope}
\end{equation}
with $\epsilon_H$ evaluated at horizon exit, $k=aH$. The factor $\tilde k^2$ arises because only
gradients of the longitudinal (Goldstone) mode carry energy. The factor $\tilde k^{-2\epsilon_H}$
follows from the amplitude $\propto H^2$ at horizon exit, since modes with larger $k$ exit later,
when $H$ is smaller. In warm inflation the radiation term dominates $\epsilon_H$: dissipation slows
the inflaton, so $\dot\phi^2/(2M_P^2H^2)\lesssim10^{-2}$, while the radiation it produces makes up
$10$--$40\%$ of the energy density during the last ten $e$-folds. Hence
$\epsilon_H\simeq0.2$--$0.8$ over this interval. Modes with $\tilde k\gtrsim1$ are still inside
the horizon at the end of inflation. Because warm
inflation passes smoothly into radiation domination, without inflaton oscillations, these modes
evolve adiabatically and are barely excited. The spectrum is therefore cut off exponentially,
$\tilde k^3n_k\simeq e^{-9\tilde k}$, rather than as a power law.

A further difference from cold inflation appears for $m_X\sim H_e$. In cold inflation these modes
become non-relativistic around the end of inflation, while the inflaton begins to oscillate. The
oscillating background imprints oscillatory features on their spectrum; see the $\tilde m=2$ curve
in Fig.~\ref{fig:cgpp_spectrum}(b). In warm inflation the strongly damped inflaton relaxes to its
minimum without oscillating, and the background evolves smoothly into radiation domination.
Consequently, the warm-inflation spectra are smooth for all masses.

\begin{figure}
    \centering
    \includegraphics[width=0.9\linewidth]{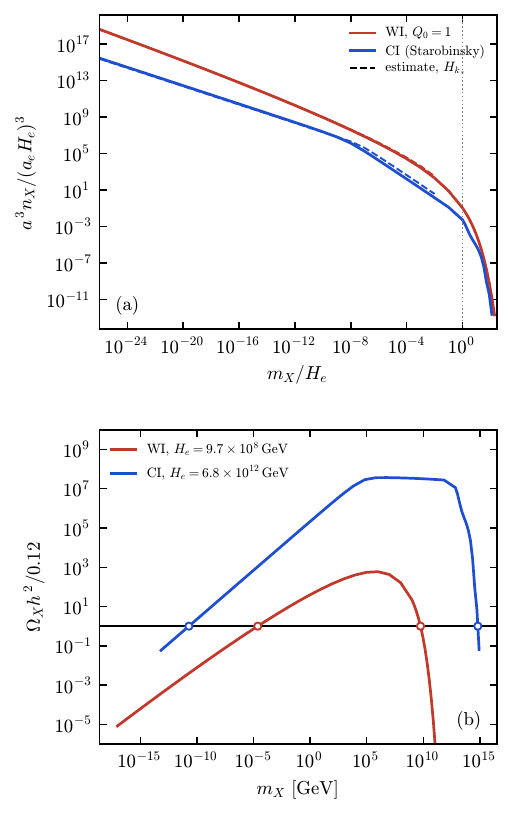}
    \caption{CGPP of a minimally coupled Proca field in warm inflation ($Q_0=1$, red) and
    Starobinsky cold inflation with $T_{\max}=T_e$ (blue). (a) Comoving number in units
    $a_e=H_e=1$; dashed lines show Eq.~\eqref{eq:nX_estimate}. (b) Relic abundance versus $m_X$,
    each model with its CMB-normalized $H_e$; open circles mark $\Omega_Xh^2=0.12$. In cold
    inflation the oscillation-averaged reheating background is used, so production by the
    oscillating inflaton is not included.}
    \label{fig:cgpp_WI_CI}
\end{figure}

To compare CGPP in warm and cold inflation, Fig.~\ref{fig:cgpp_WI_CI} shows the comoving number
for $Q_0=1$ and for Starobinsky inflation with $T_{\max}=T_e$. In units of $H_e$, warm inflation
produces $20$--$10^3$ times more light dark matter. This follows from the estimate
\begin{equation}
\frac{a^3n_X}{a_e^3H_e^3}\simeq\frac{1}{8\pi^2}\left(\frac{a_*}{a_e}\right)^{3}
\frac{m_X}{H_e}\left(\frac{H_{k_*}}{H_e}\right)^{2},
\label{eq:nX_estimate}
\end{equation}
where $H(a_*)=m_X$ and $H_{k_*}$ is the Hubble rate when $k_*=a_*m_X$ leaves the horizon. In cold
inflation $\epsilon_H\ll1$, so $H_{k_*}\simeq H_e$, whereas in warm inflation $\epsilon_H$ is of
order $0.1$--$1$ and $H_{k_*}/H_e\simeq10$--$10^3$. Equation~\eqref{eq:nX_estimate} reproduces
the numerical result to within $10\%$ in warm inflation and a factor of three in cold inflation.
In physical units, however, the larger Hubble rate of cold inflation dominates, and CGPP alone
excludes $2\times10^{-11}\lesssim m_X\lesssim7\times10^{14}$~GeV in cold inflation, compared with
$2\times10^{-5}\lesssim m_X\lesssim6\times10^{9}$~GeV in warm inflation.

\begin{figure}[t]
    \centering
    \includegraphics[width=0.8\columnwidth]{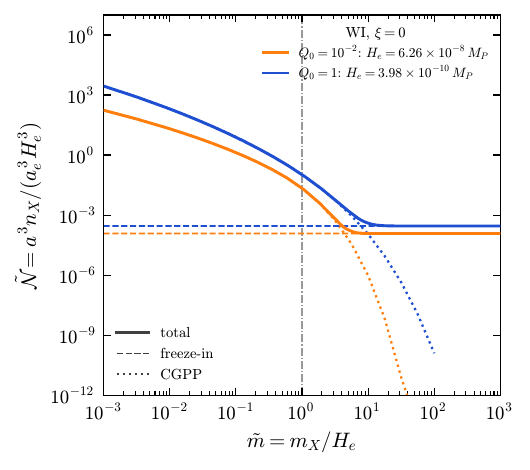}
    \caption{Comoving number $\tilde{\mathcal N}=a^3n_X/(a_e^3H_e^3)$ of purely gravitational vector
    dark matter ($\xi=0$) produced in warm inflation with $Q_0=10^{-2}$ and $1$, as a function of
    $m_X/H_e$. Dashed lines show freeze-in, dotted lines gravitational particle production, and
    solid lines the total.}
    \label{fig:number_FI_CGPP}
\end{figure}

Next, we determine which channel dominates in warm inflation for a given mass. To this end, we
evaluate the comoving number of Eq.~\eqref{eq:nX_CGPP} and compare it with that from freeze-in
computed on the same background, as shown in Fig.~\ref{fig:number_FI_CGPP} for $Q_0=10^{-2}$
and $1$. Since both channels share the background, their comoving numbers can be compared
directly. CGPP dominates for light dark matter, exceeding freeze-in by six to seven orders of
magnitude at $m_X=10^{-3}H_e$. For $m_X\gtrsim H_e$, however, it falls steeply, since such heavy
modes evolve adiabatically and are barely excited at the end of inflation. Freeze-in, in
contrast, is independent of $m_X$ for $m_X\ll T_e$, which holds over the whole range shown. The
two channels contribute equally at $m_X\simeq4.4\,H_e$ for $Q_0=10^{-2}$ and $7.8\,H_e$ for
$Q_0=1$: lighter dark matter is produced mainly by CGPP, and heavier dark matter mainly by
freeze-in from the thermal bath.

\subsection{Relic abundance from CGPP}

\begin{figure}
    \centering
    \includegraphics[width=0.8\columnwidth]{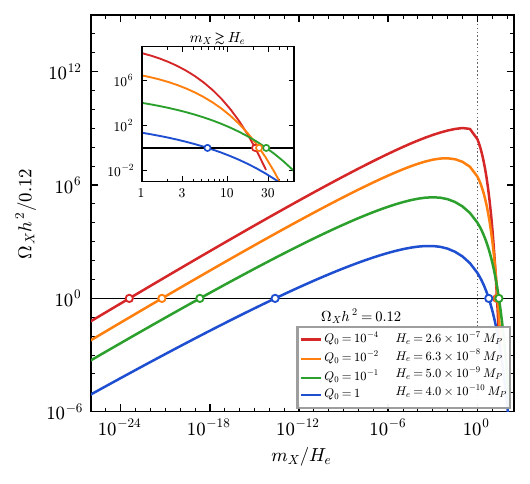}
    \caption{Relic abundance of $X$ from CGPP, $\xi=0$, $\Omega_Xh^2/0.12$, versus $m_X/H_e$ for
    $Q_0=10^{-4},10^{-2},10^{-1},1$, including both polarizations. Open circles mark
    $\Omega_Xh^2=0.12$ and the dotted line $m_X=H_e$. For $m_X\ll H_e$,
    $\Omega_X\propto m_X^{1/2-\epsilon_H}$; for $m_X\gtrsim H_e$ it is exponentially
    suppressed (inset).}
    \label{fig:cgpp_relic}
\end{figure}

Assuming that the dark matter is produced solely through CGPP, integrating the spectrum over $k$
gives the comoving number density of $X$, which we compare with the observed abundance $\Omega_Xh^2=0.12$~\cite{Planck:2018vyg}. Since $n_X/s$ is conserved
once production ends, the present abundance is
\begin{align}
\Omega_Xh^2&=\frac{m_X\,s_0h^2}{\rho_c}\,\frac{a^3n_X}{a^3s}\nonumber\\
&\simeq5.9\times10^{6}\,\frac{m_X}{\rm GeV}\left(\frac{H_e}{T_e}\right)^{3}
\frac{1}{\Delta_s}\,\frac{a^3n_X}{a_e^3H_e^3},
\label{eq:omega_cgpp}
\end{align}
where $s_0$ is the present entropy density, $\rho_c$ the critical density,
$s_e=(2\pi^2/45)g_{*s}T_e^3$, and $\Delta_s\equiv(a^3s)_f/(a^3s)_e$ accounts for the
entropy released after the end of warm inflation. The result is shown in
Fig.~\ref{fig:cgpp_relic} and Table~\ref{tab:mass_bounds_cgpp}.

\begin{table}[t]
\caption{Dark matter mass bounds from CGPP, $\Omega_Xh^2\leq0.12$, for the four warm-inflation
benchmarks with $\xi=0$, obtained from the full numerical solution.}
\label{tab:mass_bounds_cgpp}
\renewcommand{\arraystretch}{1.35}
\begin{ruledtabular}
\begin{tabular}{ccccc}
$Q_0$ & $T_e$ & $H_e$ & $m_X^{\rm light}$ & $m_X^{\rm heavy}$\\[-3pt]
 & [$10^{-5}M_P$] & [$10^{-9}M_P$] & [GeV] & [$H_e$]\\
\hline
$10^{-4}$ & $6.83$  & $264$   & $2.4\times10^{-12}$ & $21.0$\\
$10^{-2}$ & $9.83$  & $62.6$  & $9.0\times10^{-11}$ & $23.4$\\
$10^{-1}$ & $3.18$  & $4.98$  & $2.6\times10^{-9}$  & $28.4$\\
$1$       & $0.906$ & $0.398$ & $2.4\times10^{-5}$  & $5.84$\\
\end{tabular}
\end{ruledtabular}
\end{table}

We do not discuss the effect of the non-minimal coupling on CGPP in detail, since it is not the
focus of this work. Apart from the modified stability conditions of Eq.~\eqref{eq:stability},
we expect it to play a similar role as in cold inflation, and refer the reader to
Refs.~\cite{Ozsoy:2023gnl, Cembranos:2023qph, Capanelli:2024pzd, Capanelli:2024rlk,
Grzadkowski:2024oxj, Lebedev:2025ufh} for detailed studies; in Sec.~\ref{sec:relic_combined} we
nevertheless include $\xi$ in the mode equations when combining both channels.

\subsection{Relic abundance: CGPP and freeze-in}
\label{sec:relic_combined}

\begin{figure*}
    \centering
    \includegraphics[width=\textwidth]{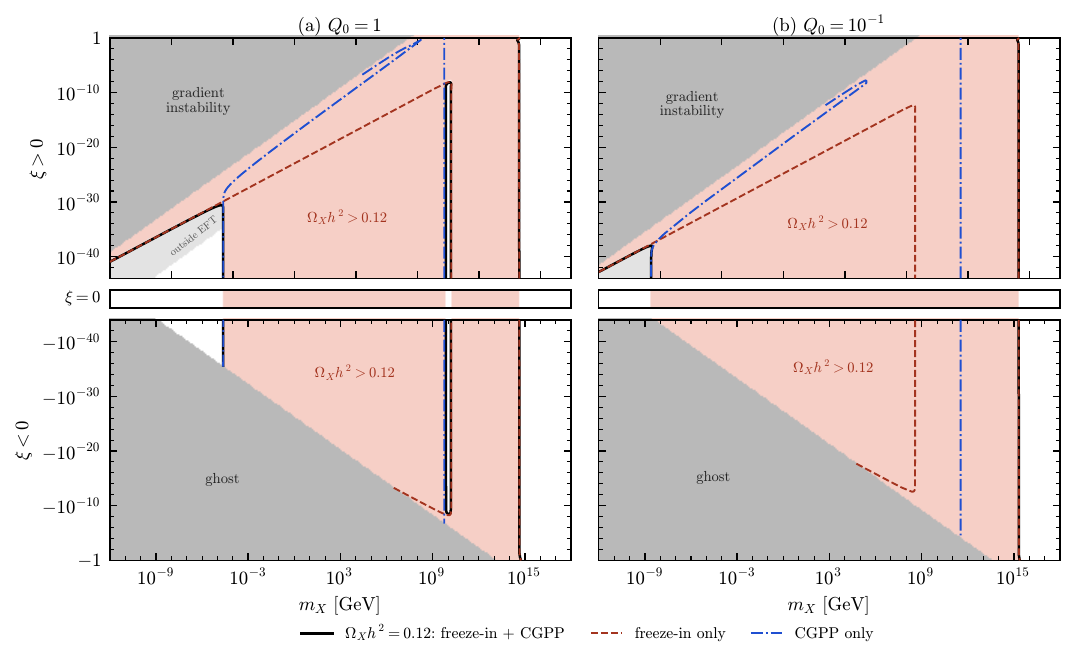}
    \caption{Constraints on the $(m_X,\xi)$ plane from freeze-in and CGPP combined, for
    (a) $Q_0=1$ and (b) $Q_0=10^{-1}$. Upper and lower panels show $\xi>0$ and $\xi<0$ on a
    logarithmic scale, and the middle strip shows $\xi=0$. The thick curve marks
    $\Omega_Xh^2=0.12$ for the total abundance, while dashed (dash-dotted) curves show freeze-in
    (CGPP) alone. Red regions are overproduced, light grey regions lie outside the EFT,
    $|\xi|\gtrsim3.5\times10^{-6}M_P^2m_X^2/T_{\max}^4$, and dark grey regions suffer from a ghost
    ($\xi<0$) or gradient ($\xi>0$) instability. White regions are allowed.}
    \label{fig:scan_combined}
\end{figure*}

Since both channels are purely gravitational, they operate simultaneously, and the relic
abundance is the sum of the two contributions,
\begin{equation}
\Omega_Xh^2=\Omega_Xh^2\big|_{\rm FI}+\Omega_Xh^2\big|_{\rm CGPP}\leq0.12 .
\label{eq:omega_total}
\end{equation}
Freeze-in from the SM bath dominates where the bath is hot and $X$ is heavy or strongly
coupled through $\xi$, whereas CGPP dominates for $m_X\lesssim H_e$, where the longitudinal mode
is efficiently excited. Figure~\ref{fig:scan_combined} shows the resulting constraints in the
$(m_X,\xi)$ plane for $Q_0=1$ and $10^{-1}$, together with the regions where the theory develops
a ghost or gradient instability [Eq.~\eqref{eq:stability}] or leaves the EFT regime.

Consider first the minimal case, $\xi=0$. Each channel overproduces dark matter only in an intermediate mass range, and the
allowed masses lie outside both ranges. For $Q_0=1$, CGPP overproduces for
$2.4\times10^{-5}\lesssim m_X\lesssim5.7\times10^{9}$~GeV and freeze-in for
$1.6\times10^{10}\lesssim m_X\lesssim4.3\times10^{14}$~GeV. For $Q_0=10^{-1}$, the corresponding
ranges are $2.6\times10^{-9}\lesssim m_X\lesssim3.4\times10^{11}$~GeV and
$3.7\times10^{8}\lesssim m_X\lesssim2.0\times10^{15}$~GeV. In both cases CGPP removes most of the
light freeze-in branch. Whether part of it survives depends on whether CGPP becomes inefficient
before freeze-in starts to overproduce. For $Q_0=1$ this is the case, and a narrow window
remains between the two excluded ranges. Since both channels contribute there, the window is
slightly narrower than the gap between them, $7.2\times10^{9}\lesssim m_X\lesssim1.5\times10^{10}$~GeV.
For $Q_0=10^{-1}$ the two ranges overlap, and no window remains.

For $\xi\neq0$ each region responds differently. On the light CGPP branch, the coupling is
restricted by the EFT bound for $\xi>0$ and by the ghost condition for $\xi<0$, to
$|\xi|\lesssim10^{-35}$ for $Q_0=1$ and $|\xi|\lesssim10^{-44}$ for $Q_0=10^{-1}$. The narrow
window for $Q_0=1$ closes at $\xi\simeq8\times10^{-9}$ ($-3\times10^{-9}$), where the $\xi^2$
term enhances freeze-in. On the heavy branch, CGPP is negligible and freeze-in depends only
weakly on $\xi$: for $Q_0=1$ the bound moves from $4.3\times10^{14}$~GeV to
$3.5\times10^{14}$~GeV at $\xi=0.5$, owing to the destructive interference discussed in
Sec.~\ref{sec:relic_freezein}, and to $5.1\times10^{14}$~GeV at $\xi=-0.5$. The heavy branch,
$m_X\gtrsim10^{14}$--$10^{15}$~GeV, is therefore the only region that remains viable for all
$|\xi|\lesssim1$.

\section{Conclusion}
\label{sec:conclusion}

In this work we studied purely gravitational production of a massive vector dark matter $X_\mu$
with a non-minimal coupling to the Einstein tensor, $\xi G_{\mu\nu}X^\mu X^\nu$, in warm
inflation. As a background we adopted $V=\lambda\phi^4$ with $\Upsilon=C_TT$, with benchmarks
consistent with the CMB constraints on $\Delta_\mathcal{R}^2$, $n_s$ and $r$
(Sec.~\ref{sec:WI} and Fig.~\ref{fig:cmb_constraints}). Expanding the metric as
$g_{\mu\nu}=\bar g_{\mu\nu}+2h_{\mu\nu}/M_P$, two unavoidable production channels emerge:
freeze-in from the SM bath through graviton exchange (Sec.~\ref{sec:grav}) and cosmological
gravitational particle production (Sec.~\ref{sec:CGPP}).

For freeze-in, we computed the scattering amplitudes $\mathrm{SM}\,\mathrm{SM}\to XX$ including,
for the first time, the contact term induced by $\xi$, which makes the cross section exactly
quadratic in $\xi$. For light $X$ the coupling enhances the rate through the longitudinal mode,
$\propto\xi^2T^4/m_X^4$, while for heavy $X$ graviton exchange and the contact term interfere
destructively, suppressing the rate by up to a factor of $50$ near $\xi\simeq0.5$
(Fig.~\ref{fig:ratio_xi}). Because the thermal bath is present throughout inflation, freeze-in is
far more efficient in warm inflation than from the SM bath in cold inflation
(Fig.~\ref{fig:yield}). From the relic abundance (Figs.~\ref{fig:relic} and~\ref{fig:relic_xi}),
we found that the light branch survives only for extremely weak couplings,
$|\xi|\lesssim10^{-8}$ for $Q_0=1$, whereas the heavy branch, $m_X\gtrsim10^{14}$--$10^{15}$~GeV,
is nearly insensitive to $\xi$ (Fig.~\ref{fig:scan_mX_xi}).

For CGPP, the smooth transition from warm inflation to radiation domination leads to spectra
without oscillatory features and with an exponential UV cutoff. Since $\epsilon_H$ is not small
during warm inflation, the peak mode leaves the horizon when $H_{k_*}\simeq10$--$10^3\,H_e$, and
the comoving number of light dark matter exceeds that in cold inflation by a factor of
$20$--$10^3$ at the same $m_X/H_e$ (Fig.~\ref{fig:cgpp_WI_CI}). Comparing the two channels on the
same background, we found that CGPP dominates for $m_X\lesssim4$--$8\,H_e$ and freeze-in for
heavier dark matter (Fig.~\ref{fig:number_FI_CGPP}); the resulting relic abundance from CGPP alone
is shown in Fig.~\ref{fig:cgpp_relic}. We also showed that the gradient instability is weaker
than in cold inflation, since the inflaton does not oscillate and thus never passes through
phases of kination, in which its kinetic energy dominates over the potential energy
($w\simeq1$), whereas the ghost condition restricts negative couplings to
$|\xi|\lesssim m_X^2/(6H_*^2)$.

Combining both channels (Fig.~\ref{fig:scan_combined}), CGPP excludes most of the light
freeze-in branch. For $Q_0=1$, dark matter is allowed for $m_X\lesssim2.4\times10^{-5}$~GeV, in a
narrow window $7.2\times10^{9}\lesssim m_X\lesssim1.5\times10^{10}$~GeV, and for
$m_X\gtrsim4.3\times10^{14}$~GeV; for $Q_0=10^{-1}$ the window closes. For $\xi\neq0$ the EFT
and the ghost condition restrict the light branch to $|\xi|\lesssim10^{-35}$, and the window
closes at $|\xi|\simeq10^{-8}$, so the heavy branch is the only region viable for all
$|\xi|\lesssim1$.

Our analysis relies on a single warm-inflation model, Maxwell--Boltzmann statistics, and
neglects thermal masses. The comparison with cold inflation does not include production by the
oscillating inflaton condensate, which would further enhance cold-inflation production. It would
be interesting to extend this study to other dissipation coefficients, such as
$\Upsilon\propto T^3$, and to more general backgrounds, for instance with a non-minimal coupling
of the inflaton to gravity, since both channels are sensitive to the expansion history. Further
extensions include the more general couplings $R\,g_{\mu\nu}X^\mu X^\nu$ and
$R_{\mu\nu}X^\mu X^\nu$, as well as the dark-matter minihalos and gravitational waves that may be
sourced by the enhanced spectrum at $k_*$.

\begin{acknowledgments}
The scientific ideas and physical reasoning in this work are the author's own. The author used
Claude Opus 5.5 (Anthropic) to assist with writing and debugging the numerical codes used for the
figures, cross-checking analytic calculations, searching the literature, and drafting and editing
the text. All calculations and AI-generated content were directed and verified by the author,
including checks against analytic limits and independent estimates, and the author takes full
responsibility for the content of this work.
\end{acknowledgments}

\section*{Data Availability}
The data that support the findings of this article were generated by numerical calculations.
The source code used to produce all figures and numerical results is openly available~\cite{code}.

\bibliographystyle{apsrev4-2}
\bibliography{my-paper}

\end{document}